%% file: main.tex
\documentclass[letterpaper,twocolumn,10pt]{article}
\usepackage{usenix}
\usepackage{cite}

\usepackage[T1]{fontenc}
\usepackage{tipa}
\usepackage{times,txfonts,color,xspace,url}
\usepackage{microtype}
\usepackage[pdftex]{graphicx}
\usepackage{subfigure}
\usepackage{rotating}
\usepackage{paralist}
\usepackage{abbrevs}
\usepackage{amssymb}
\usepackage{lastpage}
\usepackage[binary-units]{siunitx}
\DeclareGraphicsExtensions{.pdf,.png,.jpg,.eps}

\pagecolor{white}

\newcommand{\KB}[1]{\SI{#1}{\kilo\byte}}
\newcommand{\MB}[1]{\SI{#1}{\mega\byte}}
\newcommand{\GB}[1]{\SI{#1}{\giga\byte}}

\newcommand{\etal}{{\it et al.}\xspace}

\newcommand{\gm}{Green-Marl\xspace}
\newcommand{\pagerank}{PageRank\xspace}
\newcommand{\fork}{\texttt{fork()}\xspace}

\makeatletter
\renewcommand\paragraph{\@startsection{paragraph}{4}{\z@}%
                                      {2ex \@plus1ex \@minus.7ex}%
                                      {-1em}%
                                      {\normalfont\normalsize\bfseries}}
\makeatother
\newcommand{\Sys}[0]{Cichlid\xspace}
\newcommand{\shoal}[0]{Shoal\xspace}
\newcommand{\babybel}[0]{2x10 Intel Xeon E5 v2\xspace}
\newcommand{\X}[1]{\pr{x86\_{#1}}\xspace}
\makeatletter
\renewcommand\maybe@space@{%
  \maybe@ictrue % <= this is new
  \expandafter   \@tfor
    \expandafter \reserved@a
    \expandafter :%
    \expandafter =%
                 \nospacelist
                 \do \t@st@ic
  \ifmaybe@ic % <= this is new
    \space
  \fi
}
\makeatother

\newcommand{\squishlist}{
 \begin{list}{$\bullet$}
  { \setlength{\itemsep}{0pt}
     \setlength{\parsep}{3pt}
     \setlength{\topsep}{3pt}
     \setlength{\partopsep}{0pt}
     \setlength{\leftmargin}{1.5em}
     \setlength{\labelwidth}{1em}
     \setlength{\labelsep}{0.5em} } }

\newcommand{\squishend}{
  \end{list}  }

\newcommand{\pr}[1]{{\small\texttt{#1}}}

\newcommand{\linfour}{{\ttfamily 4.2.0}}
\newcommand{\lintlb}{{\ttfamily 4.2.0-tlbfs}}
\newcommand{\linthp}{{\ttfamily 4.2.0-thp}}
\newcommand{\linthree}{{\ttfamily 3.16}}
\newcommand{\lindune}{{\ttfamily 3.16-dune}}

\begin{document}

\title{\Sys: Explicit physical memory management for large
  machines}

\author{
Simon Gerber,
Gerd Zellweger,
Reto Achermann,\\
Moritz Hoffmann,
Kornilios Kourtis,
Timothy Roscoe,
Dejan Milojicic$^\dagger$\\
Systems Group, Department of Computer Science, ETH
Zurich \hspace{0.2in} {$^\dagger$Hewlett-Packard Labs}
}

%don't print date
\date{}

\maketitle

\input{abstract}
\input{introduction}
\input{linux}
\input{design}

\input{evaluation}

\input{related}

\input{conclusion}

\bibliographystyle{acm}
\bibliography{defs-abbrev,paper,l4,rialto}

\end{document}

%% file: abstract.tex
\begin{abstract}

In this paper, we rethink how an OS supports virtual memory.
Classical VM is an opaque abstraction of RAM, backed
by demand paging.  However, most systems today (from phones to
data-centers) do not page, and indeed may require the performance
benefits of non-paged physical memory, precise NUMA allocation, etc.
Moreover, MMU hardware is now useful for other 
purposes, such as detecting page access or providing large page
translation.  Accordingly, the venerable VM abstraction in OSes like
Windows and Linux has acquired a plethora of extra APIs to poke at the
policy behind the illusion of a virtual address space.

Instead, we present \Sys, a memory system which
inverts this model.  Applications explicitly manage their physical
RAM of different types, and directly (though safely) program the
translation hardware.
\Sys is implemented in Barrelfish, requires no virtualization
support, and outperforms VMM-based approaches for all but the smallest
working sets.  We show that \Sys enables use-cases for virtual memory
not possible in Linux today, and other use-cases are simple to program
and significantly faster.

\end{abstract}

%% file: introduction.tex
\section{Introduction}\label{sec:intro}

We argue that applications for modern machines should manage
physical RAM explicitly and directly program MMUs according to
their needs, rather than manipulating such hardware implicitly through a
virtual address abstraction as in Linux.  We show that explicit
primitives for managing physical memory and the MMU deliver
comparable or better application performance, greater functionality,
and a simpler and orthogonal interface that avoids the feature
interaction and performance anomalies seen in Linux. 

Traditional virtual memory (VM) systems present a conceptually simple view of
memory to the application programmer: a single, uniform virtual address space
which the OS transparently backs with physical memory.  In its pure form,
applications never see page faults, RAM allocation, address translation, TLB
misses, etc.

This simplicity has a price.  VM is an illusion --- one can exhaust physical
memory, resulting in thrashing, or the OS killing the application.  Moreover,
performance is unpredictable.   VM hardware is complex, with multiple caches,
TLBs, page sizes, NUMA nodes, etc.

For applications like databases the performance gains from
closely managing the MMU mappings and locations of physical pages on
memory controllers are as important to the end user as the functional
correctness of the
program~\cite{Giceva:2014:DQP:2735508.2735513,Leis:2014:MPN:2588555.2610507}.
Consequently, the once-simple VM 
abstraction in systems such as Linux has become steadily more 
complex, as application developers demand more control over the mapping
hardware, by piercing the VM abstraction with features like transparent huge
pages, NUMA allocation, pinned mappings, etc. In
Section~\ref{sec:linux}, we discuss the complexity, redundancy, and
feature interaction in the formerly simple VM interface. 

In response, we investigate the consequences of
turning the VM system inside-out: applications (1)
directly manage physical RAM, and (2) directly (but safely) program
MMUs to build the environment in which they
operate.  Our contribution is a comprehensive design which achieves
these goals, allows the full range of use-cases for memory system
hardware, and which performs well.

\Sys\footnote{Pronounced \textprimstress{}s\textsci{}kl\textbari{}d; see
\url{https://en.wikipedia.org/wiki/Cichlid}.}, 
a new memory management system built in the Barrelfish
research OS, adopts a radically inverted view of memory management
compared with a traditional system like Linux.  \Sys
processes still run inside a virtual address space (the MMU is enabled)
but this address space is securely constructed by the application
itself with the help of a library which exposes the full functionality of the 
MMU. Above this, all the functionality of a traditional OS memory system is
provided.

Application-level management of the virtual address space is not a new
idea; we review its history in Section~\ref{sec:related}.   \Sys
itself is an extension of the original Barrelfish physical memory management
system described in Baumann et al.~\cite{barrelfish}, which itself was
based on seL4~\cite{sel4-mm:mikes07}.  

The contributions of \Sys over these prior systems are:
\begin{itemize}

\item A comprehensive implementation of application-level memory
  management for modern hardware capable of supporting
  applications which exploit its features.  We extend the Barrelfish
  model to support safe user construction of page tables, arbitrary
  superpage mapping, demand paging, and fast access to page status
  information without needing virtualization hardware. 
\item A detailed performance evaluation of \Sys comparing it with a
  variety of techniques provided by, and different configurations of,
  a modern Linux kernel, showing that useful performance gains are
  achieved while greatly simplifying the interface.

\end{itemize}

In the next section of this paper we first review the various memory
management features in Linux as an example of the traditional
Unix-based approach.  In Section~\ref{sec:impl} we then present
\Sys, and evaluate its performance in Section~\ref{sec:eval}.
Section~\ref{sec:related} discusses the prior work on explicit
physical memory management, and Section~\ref{sec:conclusion}
summarizes the contribution and future work. 

%% file: linux.tex
\section{Background: the Linux VM system}
\label{sec:linux}

%% Discussion of Linux

We now discuss traditional VM systems, as context for \Sys.  We focus
on Unix-like systems and Linux in particular as representative of
mainstream approaches and the problems they exhibit.  Later, in
Section~\ref{sec:related} we discuss prior systems which have adopted a
different approach, some of which have strongly influenced \Sys.

\subsection{Traditional Unix}

Unix was designed when RAM was scarce, and demand paging
essential to system operation.  Virtual memory is fully decoupled
from backing storage via paging.  Each process sees a uniform
virtual address space.  All memory is paged to disk by a single system-wide
policy.  The basic virtual memory primitive visible to software is
\fork, which creates a complete copy of the virtual address space.
Modern \fork is highly optimized (e.g.\ using
copy-on-write).

Today, RAM is often plentiful, MMUs are sophisticated and
featureful devices (e.g.\ supporting superpages), and the
memory system is complex, with multiple controllers and
set-associative caches (e.g.\ which can be exploited with page
coloring).

Workloads have also changed.  High-performance multicore code
pays careful attention to locality and memory controller bandwidth.
Pinning pages is a common operation for
performance and correctness reasons, and personal devices like
phones are often designed to not page at all.   

Instead, the MMU is used for purposes aside from paging. In
addition to protection, remapping, and sharing of physical memory, MMUs are
used to interpose on main memory (e.g.\ for copy-on-write,
or virtualization) or otherwise record access (such as the
use of ``dirty'' bits in garbage collection). 

\subsection{Modern Linux}

The need to exploit the memory system fully is evident from the
range of features added to Linux over the years to ``poke
through'' the basic Unix virtual address abstraction.

The most basic of these creates additional ``shared-memory objects''
in a process' address space, which may or may not be actually shared.
Such segments are referred to by file descriptors and can either be
backed by files or ``anonymous''.  The basic operation for mapping
such an object is \texttt{mmap()}, which in addition to protection
information accepts around 16 different flags specifying whether the
mapping is shared, at a fixed address, contains pre-zeroed memory,
etc.  We describe basic usage of \texttt{mmap()} and related
calls in Section~\ref{subsec:linux:discussion}; above this are a
number of extensions.

% Anonymous mappings, and % File mappings?

% Superpages

\paragraph{Large pages:}  Modern MMUs support mappings at a coarser
granularity than individual pages, typically by terminating a
multi-level page table walk early.  For example, \X{64} supports
\MB{2} and \GB{1} \textit{superpages} as well as \KB{4}
pages, and for simplicity we assume this architecture in the
discussion that follows (others are similar). 

Linux support for superpage mappings is somewhat complex.  Firstly,
mappings can be created for large (\MB{2}) or huge (\GB{1}) pages via
a file system, \texttt{hugetlbfs}~\cite{lx:hugetlbpage,lwn:hugepages} either 
directly or through \texttt{libhugetlbfs}~\cite{lwn:libhugetlbfs}.
For each supported superpage size, a command-line argument tells the 
kernel to allocate a fixed pool of superpages at boot-time. This pool can be
dynamically resized by an administrator.  Shrinking a pool deallocates
superpages from applications using a hard-wired balancing policy. In 
addition, one superpage size is defined as a system-wide default which
will be used for allocation if not explicitly specified otherwise. 

Once an administrator has set up the page pools, users can be
authorized to create memory segments with superpage mappings, either
by mapping files created in the \texttt{hugetlbfs} file
system, or mapping anonymous segments with appropriate
flags.  Superpages may not be demand-paged~\cite{oracle:thp2}. 

The complexity of configuring different memory pools in Linux at boot
 has led to an alternative, \textit{transparent huge
  pages} (THP)~\cite{lx:transhuge,lwn:transhuge}.  When
configured, the kernel allocates large pages on page faults if
possible according to a single, system-wide policy, while a
low-priority kernel thread scans pages for opportunities to use large
pages through defragmentation.  Demand-paging is allowed by first
splitting the superpage into \KB{4} pages~\cite{oracle:thp2}.  A
typical modern \X{64} kernel is configured for
transparent support of \MB{2} pages, but not \GB{1} pages.
Alternatively, an administrator can disable system-wide THP at boot or by
writing to sysfs and programs can enable it on a per-region basis at runtime 
using\ \texttt{madvise()}. 

% NUMA

\paragraph{NUMA:} The \pr{mbind()} system call sets a
NUMA policy for a specific virtual memory region.  A policy consists
of a set of NUMA nodes and a mode: \emph{bind} to restrict allocation
to the given nodes; \emph{preferred} to prefer those nodes, but fall
back to others; \emph{interleave} allocations across the nodes, and
\emph{default} to lazily allocate backing memory on
the local node of the first thread to touch the virtual addresses.
This ``first touch'' policy has proved problematic for
performance~\cite{Dashti:2013:TMH:2451116.2451157}.
 
\pr{libNUMA} provides an additional \pr{numa\_alloc\_onnode()} call to
allocate anonymous memory on a specific node with \pr{mmap()} and
\pr{mbind()}.  Linux can move pages between 
nodes: \pr{migrate\_pages()} attempts to move all pages of a process
that reside on a set of given nodes to another set of nodes, while
\pr{move\_pages()} moves a set of pages (specified as an array of
virtual addresses) to a set of nodes.  Note that policy is expressed
in terms of virtual, not physical, memory.

There are also attempts~\cite{lwn:numasched12, lwn:autonuma12,
  lwn:numahurry12, lwn:numasched13, Dashti:2013:TMH:2451116.2451157,
  lwn:numaschedprobs14} to deal with NUMA performance issues
transparently in the kernel, by migrating threads closer to
the nodes containing memory they frequently access, or conversely migrating 
pages to threads' NUMA nodes, based on periodically revoking access to pages
and tracking usage with soft page faults.  A good generic policy,
however, may be impossible; highly performance-dependent applications
currently implement custom NUMA policies by modifying the
OS~\cite{Dashti:2013:TMH:2451116.2451157}.

% User-space page faults
\paragraph{User-space faults:}  Linux signals can be used to
reflect page faults to the application. GNU
libsigsegv~\cite{web:libsigsegv} provides a portable interface for
handling page faults: a user fault handler is called with
the faulting virtual address and must then be able to distinguish the
type of fault, and possibly map new pages to the faulting address.
When used with system calls such as \pr{mprotect()} and
\pr{madvise()}, this enables basic user-space page management.
The current limitations of this approach (both in performance and
flexibility) have led to a proposed facility for user-space demand
paging~\cite{lwn:userfault13,lwn:userfault14}.

\subsection{Discussion}
\label{subsec:linux:discussion}

Based on the simple Unix virtual address space, the Linux VM
system has evolved in response to new demands by accreting
new features and functionality.  This has succeeded up to a
point, but has resulted in a number of problems. 

%% \paragraph{Mechanism redundancy:}

\begin{figure}[t]
	\begin{center}
		\includegraphics[width=\columnwidth]{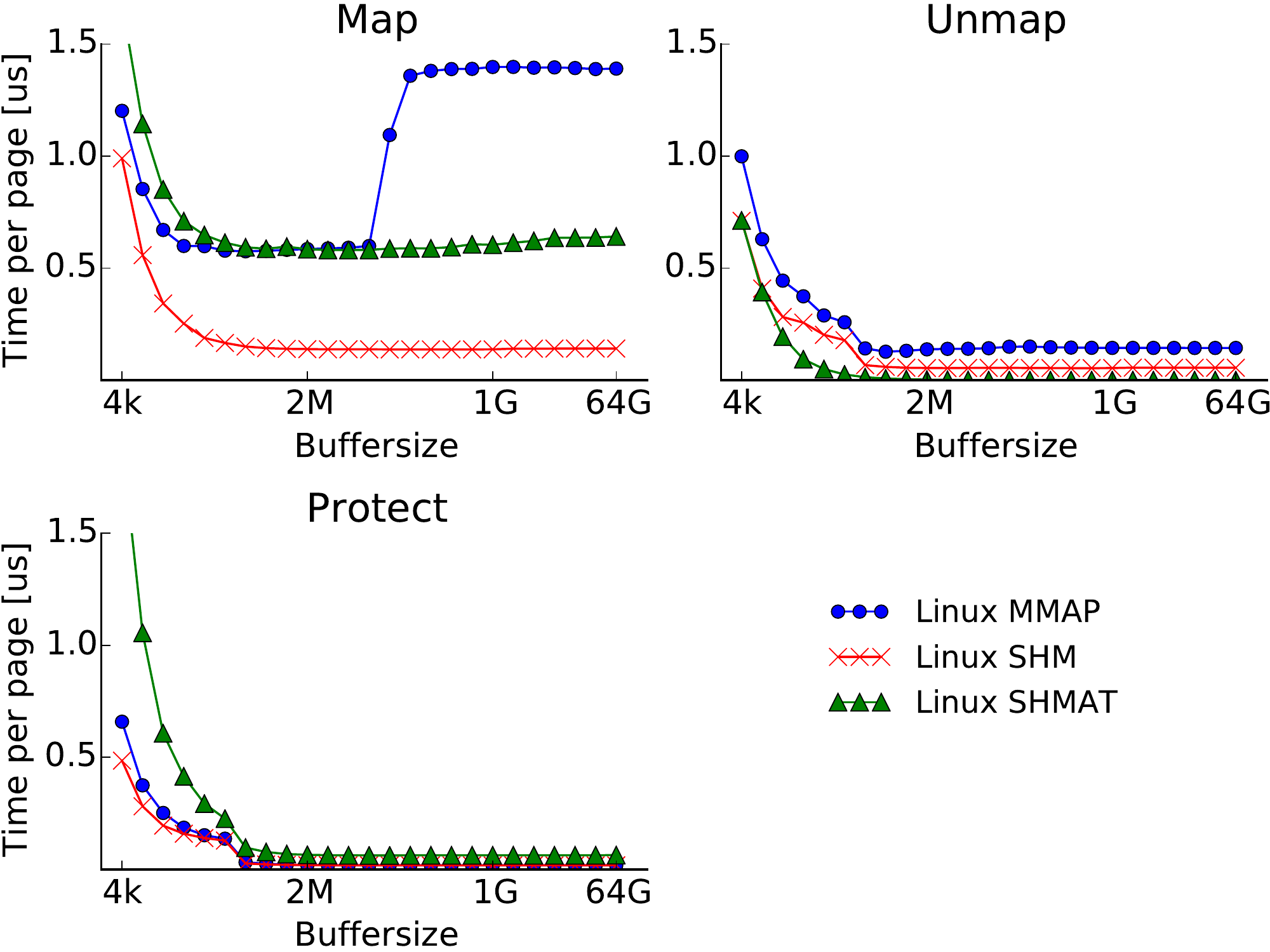}
	\end{center}
	\caption{Managing memory on Linux (\linfour)}
	\label{fig:results:memops-linux}
\end{figure}

The first is \textbf{mechanism redundancy}: there are multiple
mechanisms available to users with different
performance characteristics.   For example, 
Figure~\ref{fig:results:memops-linux} shows the performance of three
different Linux facilities for creating, destroying, and changing
``anonymous mappings'': regions of virtual address space
backed by RAM but not corresponding to a file.  
These measurements were obtained using the machine in
Table~\ref{tab:machine:babybel} using 4k pages throughout. 

%
% MMAP: http://linux.die.net/man/2/mmap ~\cite{linux:man:mmap}
% -----------------------------------------------------------------------------
%

MMAP uses an \texttt{mmap()} call with
\texttt{MAP\_POPULATE} and \texttt{MAP\_ANONYMOUS} to map and unmap
regions, and \texttt{mprotect()} for protection.  This forces
the kernel to zero pages being mapped, dominating
execution time.  Avoiding this behavior, even when safe,
requires kernel reconfiguration at build time -- a global policy
aimed at embedded systems.  
    
%
% SHM OPEN: http://linux.die.net/man/3/shm_open ~\cite{linux:man:shm}
% -----------------------------------------------------------------------------
%

SHM creates a shared memory object with
\texttt{shm\_open()} and passes it to \texttt{mmap()} and
\texttt{mprotect()}. In this case, \texttt{mmap()} will not zero the
memory.  Unmapping is also faster since memory is not immediately
reclaimed.   The object can be shared with other processes, but (unlike
MMAP mappings) cannot use large pages.
    
%
% SHMAT http://linux.die.net/man/2/shmat ~\cite{linux:man:shmat}
% -----------------------------------------------------------------------------
%

SHMAT attaches a shared segment with
\texttt{shmat()}, and \emph{does} allow large pages if the
process has the \texttt{CAP\_IPC\_LOCK} capability.  Internally, the
mechanism is similar to \texttt{mmap()}, with system-wide limits on
the number and size of segments.

%
% http://linux.die.net/man/3/get_huge_pages
% http://linux.die.net/man/7/libhugetlbfs
% http://linux.die.net/man/3/get_hugepage_region
%

For buffers up to \MB{2}, the cost per page decreases with
size for all operations due to amortization of the system call
overhead. Afterwards, the time stays constant except for MMAP map
operations. 

\texttt{libhugetlbfs} provides \texttt{get\_hugepage\_region} and 
\texttt{get\_huge\_pages} calls to directly allocate
superpage-backed memory using a \texttt{malloc}-style interface. The actual page size cannot 
be specified and depends on a system-wide default; \KB{4} pages may be 
used transparently unless the \texttt{GHR\_STRICT} flag is set. By default, 
\texttt{hugetlbfs} prefaults pages.

The high-level observation is: \emph{No single Linux API is
  always optimal, even for very simple VM operations}.  

\begin{table}
  \begin{center}
	\begin{tabular}{ll}
	        %% \hline
		CPU &  Intel Xeon
		E5-2670 v2 (Ivy Bridge)\\
		\#nodes / \#sockets / \#cores & 2 / 2 / 20 @ 2.5 GHz\\
		%core frequency & 2.4 GHz & 2.4 GHz & 2.5 GHz \\
		L1 / L2 cache & \KB{32} / \KB{256} (per core) \\
		% L1 instruction size & 64K /core & 32K /core &  32K /core\\
		L3 size &  \MB{25} (shared)\\
		dTLB (\KB{4} pages) & 64 entries (4-way) \\
		dTLB (\MB{2} pages) & 32 entries (4-way) \\
		dTLB (\GB{1} pages) & 4 entries (4-way) \\
		L2 TLB (4K) & 512 entries (4-way) \\
		RAM & \GB{256} (\GB{128} per node) \\
                Linux kernel & v.4.2.0 (Ubuntu 15.10) \\
	\end{tabular}
	\caption{Test bed specifications. \cite{intel:optref}}
	\label{tab:machine:babybel}
  \end{center}
\end{table}

\begin{table}
  \footnotesize
  \begin{center}
	\begin{tabular}{lll}
        \linfour & 4.2.0 (Ubuntu 15.10) & No large page support \\
        \lintlb & 4.2.0 (Ubuntu 15.10) & {\ttfamily hugetlbfs} enabled \\
        \linthp & 4.2.0 (Ubuntu 15.10) & Transparent huge pages enabled \\
        \linthree & 3.16 & Stock 3.16 kernel \\
        \lindune & 3.16 & Linux 3.16 with Dune \\
	\end{tabular}
	\caption{Tested Linux configurations}
	\label{tab:linux:conf}
  \end{center}
\end{table}

A second problem is \textbf{policy inflexibility}.  While the
appropriate policy for many memory management operations such as page
replacement, NUMA allocation or handling of superpages depend
strongly on individual application's workloads. In Linux, however, they usually 
either apply system-wide, require administrator configuration (often at boot), 
must be enabled at compile time, or a combination of them. 

For example, supporting two superpage sizes in \pr{hugetlbfs} requires
two different, pre-allocated pools of physical memory, each assigned
to a different file system, precluding a dynamic algorithm that could
adapt to changing workloads. 

% Problems with transparent huge page support

In addition to the added complexity in
the kernel~\cite{lwn:thp_issues}, the system-wide policies in
\emph{transparent} superpage support have led to a variety of
performance issues: Oracle DB has suffered from I/O performance
degradation when reading large extents from
disk~\cite{oracle:thp,oracle:thp2}. Redis incurs unexpected latency 
spikes using THP due to copy-on-write overhead for large pages, since
the application periodically uses \texttt{fork()} to persist database
snapshots~\cite{redis:thp}.  The \pr{jemalloc} memory
allocator experiences performance anomalies due to its use of
\pr{madvise} to release small regions of memory inside of bigger
chunks which have been transparently backed by large 
pages --- the resulting holes preventing later merging of the region
back into a large page~\cite{jemalloc:thp}. 

These issues are not minor implementation bugs, but arise from
the philosophy that memory system complexity should
be hidden from applications, and resource allocation policies should
be handled transparently by the kernel. 

The third class of problem is \textbf{feature interaction}.  We have
seen how superpages cannot be demand paged (even though modern
SSDs can transfer 2MB pages with low latency). 
Another example is the complex and subtle interaction between 
kernel-wide policies for NUMA allocation 
with superpage support~\cite{lx:hugetlbpage}.   At one level, this shows
up in the inability to control initial superpage allocation at boot
time (superpages are always balanced over all NUMA nodes).   Worse,
Gaud et al.~\cite{Gaud:2014:LPM:2643634.2643659} show that treating
large pages and NUMA separately does not work well: 
large pages hurt the performance of parallel applications on NUMA
machines because \emph{hot pages} are more likely, and larger, and
\emph{false page sharing} makes replication or migration less
effective.  Accordingly, the
Carrefour~\cite{Dashti:2013:TMH:2451116.2451157} system modifies the
kernel's NUMA-aware page placement to realize its performance gains.
\vspace{0.1in}

Collectively, these issues motivate investigating alternative
approaches.  As memory hardware diversifies in the future, memory
management policies will become increasingly complicated.  We note
that none of the Linux memory APIs actually deal with \emph{physical}
memory directly, but instead select from a limited number of complex,
in-kernel policies for backing traditional \emph{virtual} memory.

In contrast, therefore, \Sys safely exposes to programs and runtime
systems both physical memory and translation hardware, and allows
libraries to build familiar virtual memory abstractions above this.

%% file: design.tex
\section{Design}\label{sec:impl}

We now describe the design of \Sys, and how it is implemented over
the basic memory functionality of Barrelfish. 
While \Sys allows great flexibility in arranging an address space, it
nevertheless ensures the following key safety property: \emph{no \Sys
  process can issue read or write instructions for any area of
  physical memory for which it does not have explicit access
  rights.}  

Subject to this requirement, \Sys also provides the following
completeness property: \emph{a \Sys process can create any address
  space layout permitted by the MMU for which it has
  sufficient resources}.  In other words, \Sys itself poses no
restriction on how the memory hardware can be used.

There are three main challenges in the implementation that \Sys must
address: Firstly, it must securely name and authorize access to,
and control over, regions of physical memory.  \Sys achieves this using
\emph{partitioned capabilities}.  Secondly, it must allow safe
control of hardware data structures (such as page tables) by
application programs.  This, is achieved by considerably extending the
set of memory types supported by the capability system in Barrelfish
(and seL4) for \Sys to use.  Finally, \Sys must give applications
direct access to information provided by the MMU (such as access and
write-tracking bits in the page tables).  Unlike prior approaches
which rely on virtualization technology, \Sys allows direct read-only
access to page table entries; we explain below why this is safe. 

\Sys has three main components: First, the kernel 
provides capability invocations that allow application
processes to install, modify and remove page table entries and query for
the base address and size of physical regions.  Second, the kernel
exception handler redirects any exceptions generated by the MMU to the
application process that caused the exception.  Thirdly, a runtime
library provides to applications an abstraction layer over the capability system
which exposes a simple, but expressive API for managing page tables. 

% % % % % % % % % % % % % % % % % % % % % % % % % % % % % % % % % % % % % % % 
\subsection{Physical memory allocation}

\Sys applications directly allocate regions of physical memory and
pass around authorization for these regions in the form of
capabilities.  Regions can be mapped into a virtual address space by
changing a page table, or used for other purposes such as
holding page tables themselves.

\Sys extends the Barrelfish capability design, itself
inspired by seL4~\cite{sel4:iies08,sel4:sosp09,sel4:refman}.  All
physical regions are represented by capabilities, which also
confer a particular \emph{memory type}.  For example,
the integrity of the capability system itself is ensured by storing
capability representations in memory regions of type \texttt{CNode},
which can never be directly written by user-space programs.  Instead, a
region must be of type \texttt{Frame} to be mapped writable into a
virtual address space. Holding both \texttt{Frame} and \texttt{CNode}
capabilities to the same region would enable a process to forge new
capabilities by directly manipulating their bit representations, and
so is forbidden. Such a situation is prevented by having a kernel
enforced type hierarchy for capabilities.

Capabilities to memory regions can be split and \emph{retyped}
according to a set of rules.  At system start-up, all memory is
initially of type \texttt{Untyped}, and physical memory is allocated
to processes by splitting the initial untyped region. Retyping and other 
operations on capabilities is performed by system calls to the kernel. 

seL4 capabilities are motivated by the desire to prove correctness
properties of the seL4 kernel, in particular, the property that no
system call can fail due to lack of memory. Hence, seL4 and Barrelfish  perform 
no dynamic memory allocation in the kernel, instead memory for all dynamic 
kernel data structures is allocated by user-space programs and retyped
appropriately, such as to a kernel thread control block or a
\texttt{CNode}, for example.

Capabilities are attractive since they export physical
memory to applications in a safe manner: application may not
arbitrarily use physical memory; they must instead ``own'' the corresponding
capability. Furthermore, capabilities can be passed
between applications.  Finally, capabilities have some characteristics
of objects: each capability type has a set of \emph{operations} which
can be invoked on it by a system call.

In Barrelfish, seL4, and \Sys, the kernel enforces safety using two
types of meta-data: a \emph{derivation database} and a per-processes
\emph{capability space}.  All capability objects managed by a kernel
are organized in a capability derivation tree. This tree enables
efficient queries for descendants (of retype and split operations) and
copies.  These queries are used to prevent retype races on separate
copies of a capability that might compromise the system.

User processes refer to capabilities and invoke operations on
them using opaque handles.  
Each process has its own capability
address space, which is explicitly maintained via a radix tree in the
kernel which functions as a \emph{guarded page table}. The nodes of
the tree are also capabilities (retyped from RAM capabilities) and are
allocated by the application.  

The root of the radix tree for each process is stored in the process
control block.  When a process invokes a capability operation it
passes to the kernel the capability handle with the invocation
arguments.  To perform the operation, the kernel traverses the process'
capability space to locate the capability corresponding to the handle
and authorizes the invocation. 

\Sys builds on the basic Barrelfish capability mechanisms to allow
explicit allocation of different kinds of memory. A memory region has
architectural attributes such as the memory controller it resides on,
whether it is on an external co-processor like a GPGPU or Intel Xeon
Phi, whether it is persistent, etc.  Applications explicitly acquire
memory with particular attributes by requesting a capability from an
appropriate memory allocator process, of which there are many.
Furthermore, less explicit ``best effort'' policies can be layered on
top by implementing further virtual allocators which can, for example,
steal RAM from nearby controllers if local memory is scarce.

% % % % % % % % % % % % % % % % % % % % % % % % % % % % % % % % % % % % % % % 
\subsection{Securely building page tables}\label{subsec:pagetables}

Page tables are hardware specific, and at the lowest level, \Sys's
interface (like seL4 and Barrelfish) reflects the actual hardware.
Applications may use this interface directly, or a high-level API with
common abstractions for different MMUs, to safely build page tables,
exchange page tables on a core, and install mappings for any physical
memory regions for which the application is authorized.  The choice of
virtual memory layout, and its representation in page tables, is fully
controlled by the application.  Cores can share sub-page-tables
between different page-table hierarchies to alias a region of memory
at a different address or to share memory between different cores as
in Corey~\cite{Boyd-Wickizer:2008:COS:1855741.1855745}.

\Sys adds support for multiple page sizes (\MB{2} and \GB{1}
superpages in \X{64}, and \MB{16}, \MB{1}, and \KB{64} pages in
ARMv7-a~\cite{armv7}) to the Barrelfish memory management
system~\cite{barrelfish}. 
\Sys decouples the physical memory allocation from programming the MMU.
Therefore the API allows for a clean way to explicitly select the page size
for individual mappings, map pages from a mixture of different 
page sizes, and change the virtual page sizes for 
mappings of contiguous physical memory regions all directly
from the applications itself instead of relying on the kernel to
implement the correct policy for all cases.

To do this, \Sys extends the Barrelfish memory system
(and that of seL4) by introducing a new capability type for every level
of page table for every architecture supported by the OS.  This is
facilitated by the \emph{Hamlet} domain-specific language for
specifying capability types~\cite{dagand:fof:plos09}.

For example, for an MMU in \X{64} long-mode there are four different
types of page table capability, corresponding to the 4 levels of a
64-bit x86 page table (\texttt{PML4}, \texttt{PDPT}, \texttt{PD}, and
\texttt{PT}).   A \texttt{PT} (last-level page table) capability can
only refer to a 4k page-aligned region of RAM and has a \texttt{map}
operation which takes an additional capability plus an entry number as
arguments.  This capability in turn must be of type \texttt{Frame}
and refer to another 4k page.  The operation installs the appropriate
page table entry in the \texttt{PT} to map the specified frame.  
The kernel imposes no policy on this mapping, other than restricting
the type and size of capabilities. 

Similarly, a \texttt{map} on a \texttt{PD} (a 2nd-level ``page
directory'') capability only accepts a capability argument which is of
size \KB{4} and type \texttt{PT}, \emph{or} of type
\texttt{Frame} and size \MB{2} (signifying a large page mapping). 

A small set of rules therefore captures all possible valid and
authorized page table operations for a process, while excluding any
that would violate the safety property.  Moreover, checking these
rules is fast and is partly responsible for \Sys's superior
performance described in Section~\ref{sec:eval:memops:compare}.  This
type system allows user-space \Sys programs to construct flexible page
tables while enforcing the safety property stated at the start of this
section. 

\newcommand{\map}{\texttt{map}} \newcommand{\unmap}{\texttt{unmap}}
\newcommand{\modifyflags}{\texttt{modify\_flags}}
\newcommand{\clear}{\texttt{clear\_dirty\_bits}}
\newcommand{\identify}{\texttt{identify}} \Sys's full kernel interface
contains the following capability invocations: \identify, \map,
\unmap, \modifyflags{} (protect), and \clear.

Memory regions represented by capabilities and associated rights
allow user-level applications to safely construct page tables; they
allocate physical memory regions and retype them to hold a page
table and install the entries as needed. 

Typed capabilities ensure a process cannot
successfully map a physical region for which it does not have authorization.
The process of mapping itself is still a privileged operation handled
by the kernel, but the kernel must only validate the references and
capability types before installing the mapping. Safety is guaranteed
based on the type system: page tables have a specific type which
cannot be mapped writable.

Care must be taken in \Sys to handle capability revocation.  In
particular, when a \texttt{Frame} capability is revoked, all page
table entries for that frame must be quickly identified and removed.
\Sys handles this by requiring each instance of a \texttt{Frame}
capability to correspond to at most one hardware page table entry.  To
map a frame into multiple page tables, or at multiple locations in the
same page table, the program must explicitly create copies of the
capability.

As described so far, each operation requires a separate
system call.  \Sys optimizes this in a straightforward way by allowing
batching of requests, amortizing system call cost for large region
operations.  The \map, \unmap, and \modifyflags{} operations all take
multiple consecutive entries for a given page table as arguments.

In Section~\ref{sec:eval:gups} we confirm existing work on the effect
of page size on performance of particular workloads, and in
Section~\ref{sec:eval:heteropages} we show that the choice of the page
size is highly dynamic and depends on the program's configuration such
as the number of threads and where memory is allocated.

In contrast, having the OS transparently select a page size is an
old idea~\cite{Navarro:2002:PTO:1060289.1060299} and is the default
in many Linux distributions today, but finding a policy that
satisfies a diverse set of different workloads is difficult in
practice and leads to inherent complexity with questionable
performance
benefits~\cite{Gaud:2014:LPM:2643634.2643659,Gorman:2010:PCE:2185870.2185899,oracle:thp,redis:thp}.

% % % % % % % % % % % % % % % % % % % % % % % % % % % % % % % % % % % % % % % 
\subsection{Page faults and access to status bits}

\Sys uses the existing Barrelfish functionality for reflecting
VM-related processor exceptions back to the faulting process, as in
Nemesis~\cite{Hand:1999:SNO:296806.296812} and
K42~\cite{k42:eurosys06}.  This incurs lower kernel overhead than
classical VM and allows the application to implement its own paging
policies. In Sections~\ref{sec:eval:appel} and \ref{sec:eval:micro} we
show that \Sys's trap latency to user space is considerably lower than
in Linux.

\Sys extends Barrelfish to allow page-traps to be eliminated for
some use-cases when the MMU maintains page access information in the
page table entries.  While Dune~\cite{Belay_dune} uses nested paging
hardware to present ``dirty'' and ``accessed'' bits in an \X64 page
table to a user space program, \Sys achieves this
\emph{without} hardware support for virtualization.

We extend the kernel's mapping rules in
Section~\ref{subsec:pagetables} to allow page tables themselves to be
mapped read-only into a process' address space.  Essentially, this
boils down to allowing a \KB{4} capability of type \texttt{PML4},
\texttt{PDPT}, \texttt{PD}, or \texttt{PT} to be mapped in an entry in
a \texttt{PT} instead of a \texttt{Frame} capability, with the added
restriction that the mapping must be read-only.  

This allows applications (or libraries) to read ``dirty'' and
``accessed'' bits directly from page table entries without trapping to
the kernel.  Setting or clearing these bits remains a privileged
operation which can only be performed by a kernel invocation passing
the capability for the page table.

Note that this functionality remains safe under the capability system:
an application can only access the mappings it has installed itself
(or for which it holds a valid capability), and cannot subvert them. 

In Section~\ref{sec:eval:boehm} we demonstrate the benefits of
this approach for a garbage collector.
\Sys's demand-paging functionality for \X{64} also uses mapped dirty
bits to determine if a frame's contents should be paged-out before
reusing the frame.  

Since \Sys doesn't need hardware virtualization support, such
hardware, if present, can be used for virtualization. \Sys can work
both inside a virtual machine, or as a better memory management system
for a low-level hypervisor.

Moreover, nested paging has a performance cost for large working sets,
since TLB misses can be twice as expensive.  In
Section~\ref{sec:eval:nestedgups} we show that for small working sets
(below \MB{16} for our hardware) a Dune-like approach outperforms \Sys
due to lower overhead in clearing page table bits, but for
medium-to-large working sets \Sys's lower TLB miss latency improves
performance.  

The \Sys and Dune approaches are complementary, and a natural
extension to \Sys (not pursued here) would allow applications access
to both the physical (machine) page tables \emph{and} nested page
tables if the workload can exploit them.

% % % % % % % % % % % % % % % % % % % % % % % % % % % % % % % % % % % % % % % 
\subsection{Runtime library}

\Sys provides a number of APIs above the capability invocations
discussed above. 

The first layer of \Sys's user-space wraps the invocations to
create, modify or remove single mappings, and keep track of
the application's virtual address space layout. 

While application programmers can build directly on this, the \Sys
library provides higher-level abstractions based on the concepts of
\emph{virtual regions} (contiguous sets of virtual addresses), and
\emph{memory objects} that can be used to back one or more virtual
regions and can themselves be comprised of one or more physical
regions.

This layer is important to \Sys's usability.  Manually invoking
operations on capabilities to manage the virtual address space can be
cumbersome; take the example of a common operation such as
mapping an arbitrarily-sized region of physical memory $R$ with physical base
address $P$ and size $S$ bytes, $R = (P, S)$, at
an arbitrary virtual base address $V$.  The number of invocations needed
to create this simple mapping varies based on $V$, $S$, and the desired
properties of the mapping (such as page size), as well as the state of
the application's virtual address space before the operation.  In
particular, installing a mapping can potentially entail creating
multiple levels of page table in addition to installing a page table
entry.  The library encapsulates the code to do this on demand, as well
as batching operations up to amortize system call overhead. 

Finally, the library also provides traditional interfaces such
as \pr{sbrk()} and \pr{malloc()} for areas of memory where performance
is not critical.  To simplify start-up, programs running
over \Sys start up with a limited, conventional virtual address space
with key segments (text, data, bss) backed with RAM, though this
address space is, itself, constructed by the process' parent using
\Sys (rather than the kernel).   

In addition, the \Sys library provides demand paging to disk as in
Nemesis~\cite{Hand:1999:SNO:296806.296812}, but not by default: many
time-sensitive applications rely on \emph{not} paging for correctness,
small machines such as phones typically do not page anyway, and and
the growth of non-volatile main
memory~\cite{hotos:themachine,themachine_website} may make
demand-paging obsolete.  Unlike the Linux VM system, demand paging is
orthogonal to page size: the \Sys library can demand-page superpages
provided the application has sufficient frames and disk space. Furthermore, the 
application is aware of the number of backing frames and can add or remove  
frames explicitly at runtime if required.

The library shows that building a classic VM abstraction over 
\Sys is straightforward, but the reverse is not the case.

%% file: evaluation.tex
\section{Evaluation}\label{sec:eval}

We evaluate \Sys by first demonstrating that primitive operations have
performance as good as, or better than those of Linux, and then showing that
\Sys's flexible interface allows application programmers to usefully optimize
their systems.  

All Linux results, other than those for Dune (Section~\ref{sec:eval:boehm}),
are for version 4.2.0, as shipped with Ubuntu 15.10, with three large-page
setups: none, \pr{hugetlbfs}, and transparent huge pages.  As the Dune patches
(git revision 6c12ba0) require a version 3 kernel, these benchmarks use kernel
version 3.16 instead.  These configurations are summarized in
Table~\ref{tab:linux:conf}.  Thread and memory pinning was done using
\pr{numactl} and \pr{taskctl}.  Performance numbers for Linux are always the
best among all tested configurations.

\subsection{Appel and Li benchmark}\label{sec:eval:appel}

\begin{figure}[t]
    \begin{center}
        \includegraphics[page=1,width=\columnwidth]{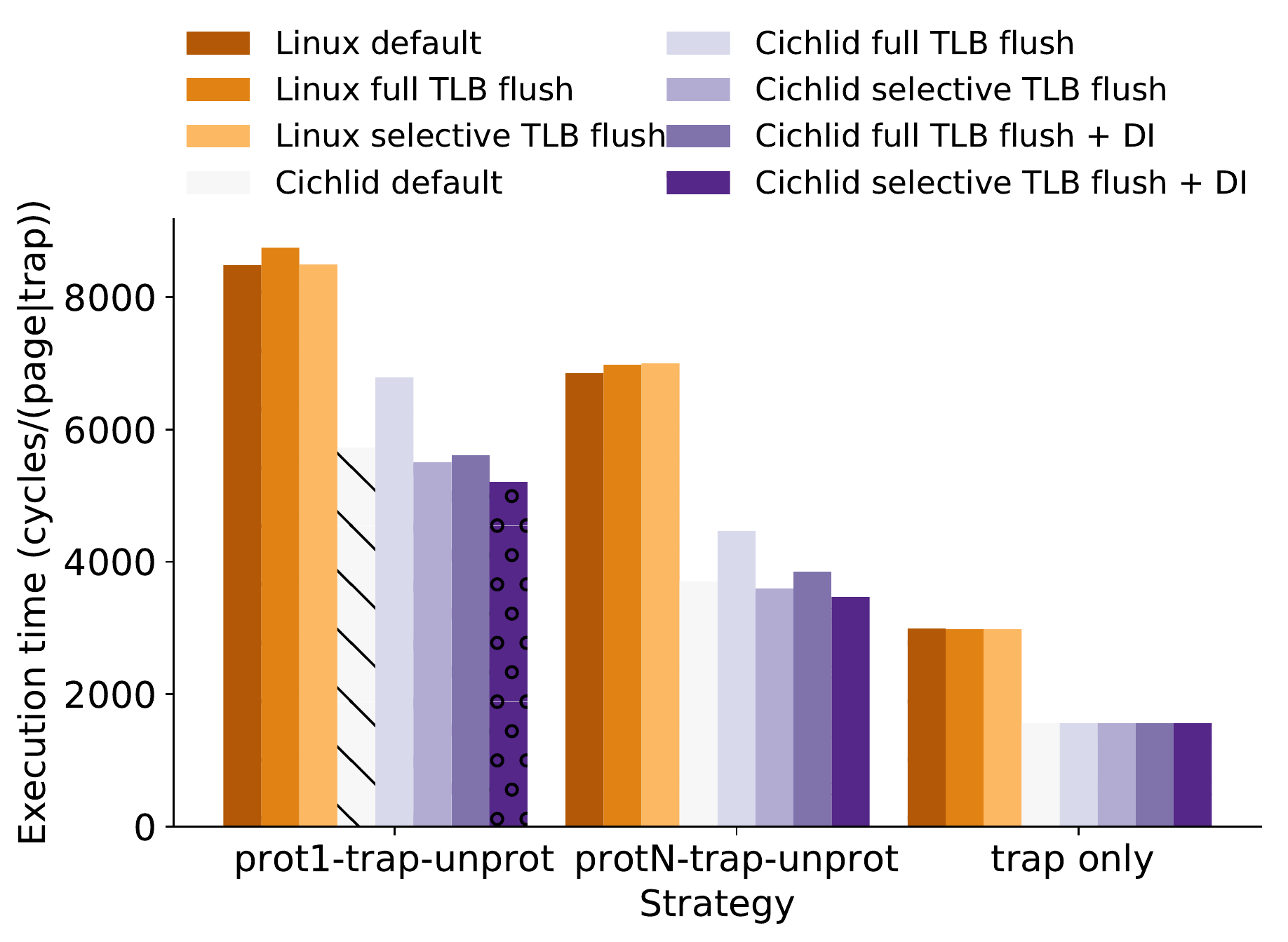}
    \end{center}
    \caption{Appel-Li benchmark. (Linux \linfour)}
    \label{fig:appel-li}
\end{figure}

The Appel and Li benchmark~\cite{Appel:1991:VMP:106972.106984} tests
operations relevant to garbage collection and other non-paging tasks.  This
benchmark is compiled with flags \pr{-O2 -DNDEBUG}, and summarized in
Figure~\ref{fig:appel-li}. 

We compare Linux and \Sys with three different TLB flush modes: 1) Full:
Invalidate the whole TLB (writing \pr{cr3} on \pr{x86}) every time, 2)
Selective: Only invalidate those entries relevant to the previous
operation (using the \pr{invlpg} instruction), and 3) System default:
\Sys, by default, does a full flush only for more than one page.  Linux's
default behavior depends on kernel version. The version tested (4.2.0) does a
selective flush for up to 33 pages, and full a flush
otherwise~\cite{lx:x86tlbflush}.  We vary this value to change Linux's flush
mode.  The working set here is less than \MB{2}, and thus large pages have no
effect and are disabled.

\Sys is consistently faster than Linux here.

For multi-page protect-trap-unprotect (\emph{protN-trap-unprot}), \Sys is 46\%
faster than Linux.  For both systems, the default adaptive behavior is as
good as, or better than, selective flushing.  The \Sys\emph{+DI} results use the
kernel primitives directly, to isolate the cost of user-space accounting,
which is around 5\%.

%
% operation microbenchmarks input
% ----------------------------------------------------------------------------
\input{evaluation-memops}

%
% GUPS
% ----------------------------------------------------------------------------

\subsection{Random accesses benchmark (GUPS)}
\label{sec:eval:gups}

Many HPC workloads have a random memory access pattern, and spend up to 50\%
of their time in TLB misses~\cite{Soma:2014:RVM:2612262.2612264}.  Using the
RandomAccess benchmark~\cite{gups_randomaccess} from the HPC
Challenge~\cite{gups} suite, we demonstrate that carefully user-selected page
sizes, as enabled by \Sys, have a dramatic performance effect.

We measure update rate (Giga updates per second, or GUPS) for
read-modify-write on an array of 64-bit integers, using a single thread.  We
measure working sets up to \GB{32}, which exceeds TLB coverage for all page
sizes. Linux configuration is \lintlb, with pages allocated from the local
NUMA node. If run with transparent huge pages instead, the system always
selects \MB{2} pages, and achieves lower performance.

\begin{table}
	\begin{center}
		{\small
			\begin{tabular}{c|rr|rr}
				& \multicolumn{2}{c}{\Sys} & \multicolumn{2}{|c}{Linux} \\
				Page Size & GUPS & Time & GUPS & Time \\
				\hline
				4k &  0.0122 & 1397s & 0.0121 & 1414s \\
				%4k & LCG &  0.0122 & (1396s) & 0.0123 & (1396s) \\
				2M &  0.0408 & 420s  & 0.0408 & 421s  \\
				%2M & LCG &  0.0411 & (418s)  & 0.0409 & (419s)  \\
				1G &  0.0659 & 260s  & 0.0658 & 261s  \\
				%1G & LCG &  0.0672 & (255s)  & 0.0668 & (257s)  \\
			\end{tabular}
		}
	\end{center}
	\caption{GUPS as a function of page size, \GB{32} table.}
	\label{tab:results:gups}
\end{table}

\begin{figure}[t]
	\begin{center}
		\includegraphics[width=\columnwidth]{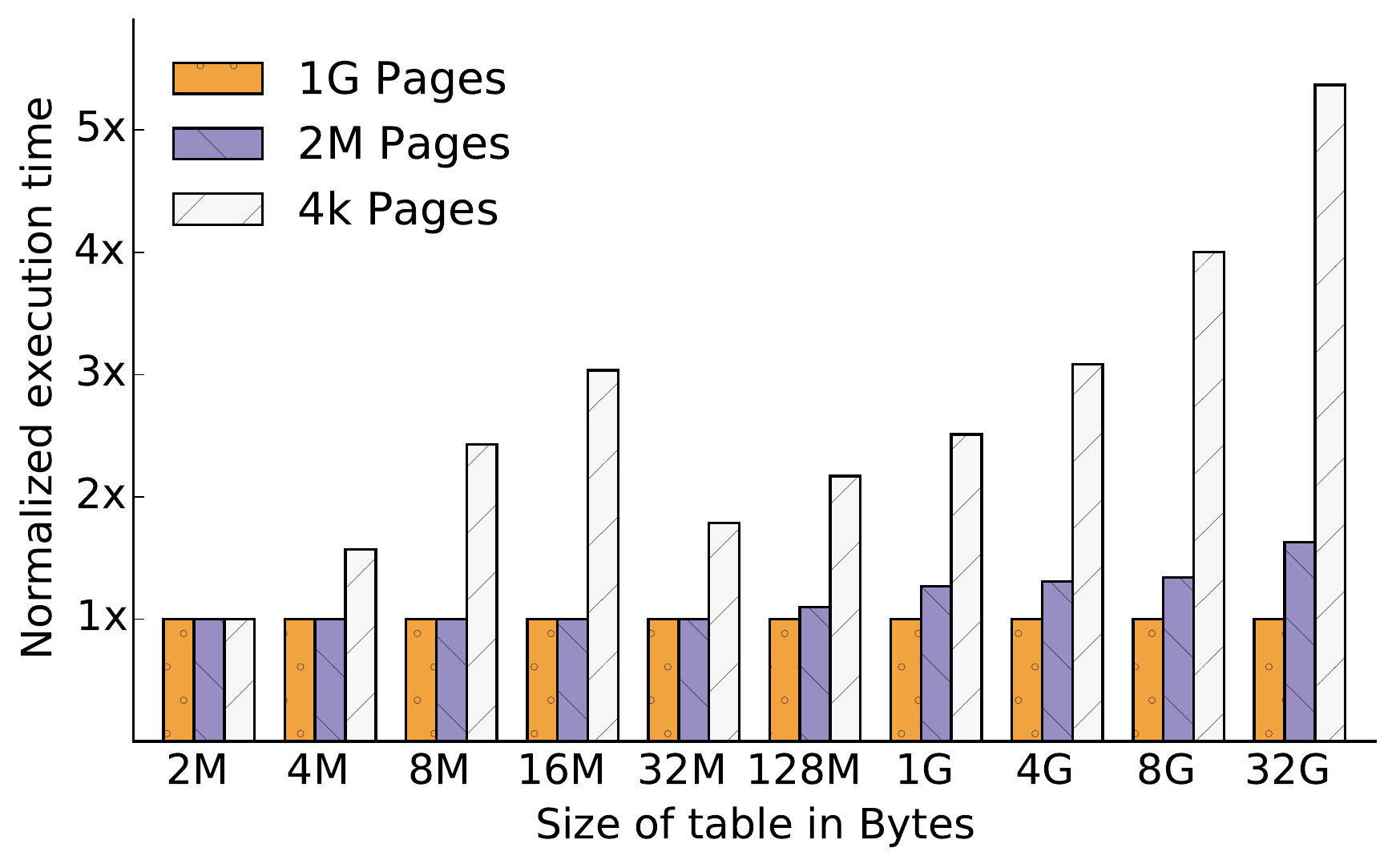}
	\end{center}
	\caption{GUPS as a function of table size, normalized.}
	\label{fig:results:gups-barrelfish-normalized}
\end{figure}

Figure~\ref{fig:results:gups-barrelfish-normalized} shows the results on \Sys,
normalized to \GB{1} pages.  Performance drops once we exceed TLB coverage: at
\MB{2} for \KB{4} pages, and at \MB{128} for \MB{2} pages.  The apparent
improvement at \MB{32} is due to exhausting the L3 cache, which slows all three
equally, bringing the normalized results together.  Large pages not only
increase TLB coverage, but cause fewer table walk steps to service a TLB miss.
Page-structure caches would reduce the number of memory accesses even
further but are rather
small~\cite{Bhattacharjee:2013:LMM:2540708.2540741,Barr:2010:TCS:1815961.1815970}
in size.  \Sys and Linux perform identically in the test, as
Table~\ref{tab:results:gups} shows.  These results support previous findings
on TLB overhead~\cite{Soma:2014:RVM:2612262.2612264,
Basu:2013:EVM:2485922.2485943}, and emphasize the importance for applications
being able to select the correct page size for their workload.

\begin{figure}[t]
    \begin{center}
        \includegraphics[width=\columnwidth]{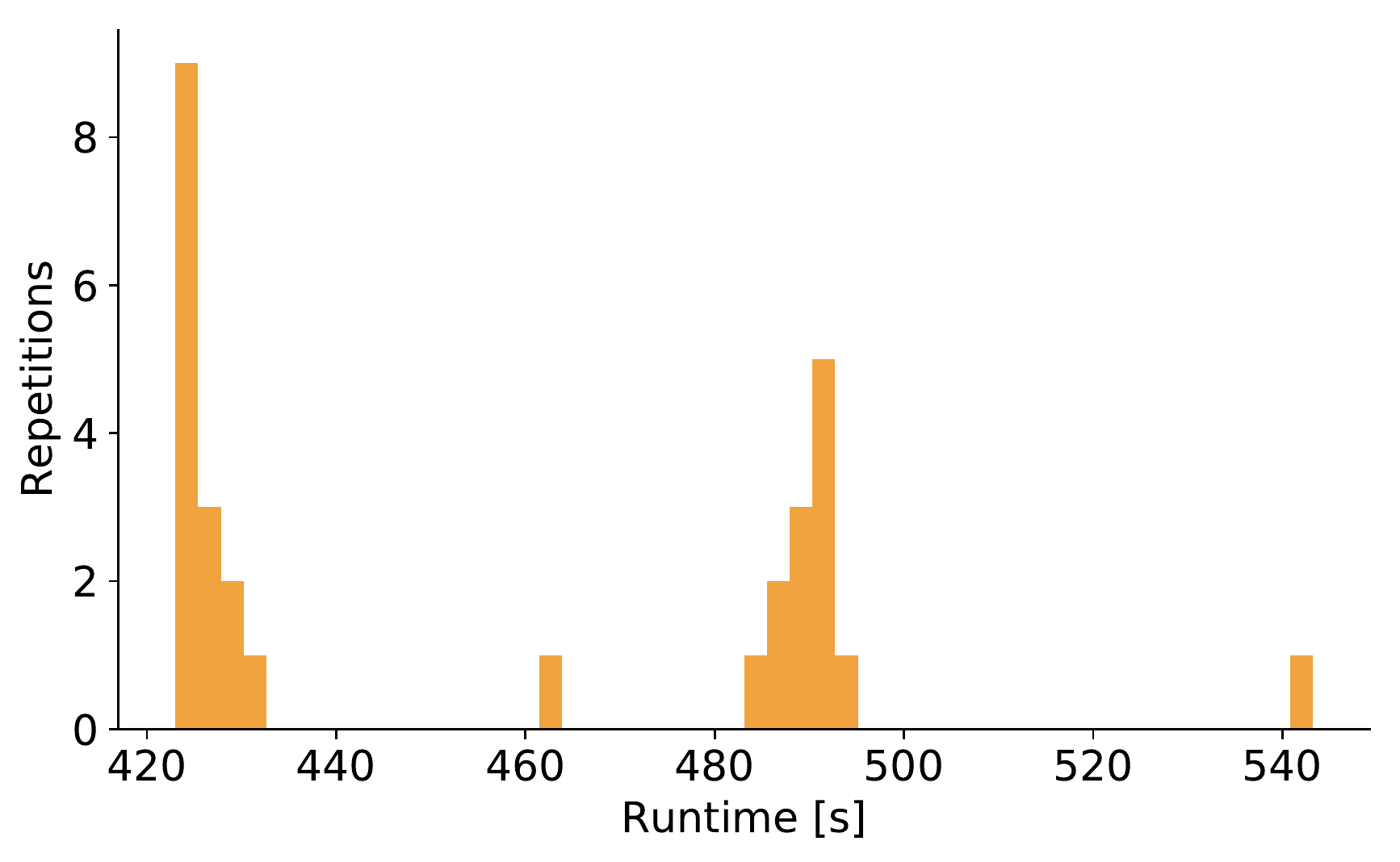}
    \end{center}
    \caption{GUPS variance. \lintlb, \MB{2} pages.}
    \label{fig:results:gups-hist}
\end{figure}

On Linux, even with NUMA-local memory, high scheduling priority, and no
frequency scaling or power management, there is a significant variance between
benchmark runs, evidenced by the multimodal distribution in
Figure~\ref{fig:results:gups-hist}.  This occurs for both \pr{hugetlbfs} and
transparent huge pages, and is probably due to variations in memory
allocation, although we have been unable to isolate the precise cause.  This
variance is completely absent under \Sys even when truly randomizing
paging layout and access patterns, demonstrating again the benefit of
predictable application-driven allocation.

\subsection{Mixed page sizes}\label{sec:eval:heteropages}

Previous work~\cite{Gaud:2014:LPM:2643634.2643659} has shown that while large
pages can be beneficial on NUMA systems, they can also hurt performance.
Things are even more complicated when there are more page sizes (e.g., \KB{4},
\MB{2}, \GB{1} for \X{64}). Furthermore, modern machines often have a distinct
TLB for each page size, suggesting that using a mix of page sizes increases
TLB coverage.

Kaestle~\etal~\cite{Kaestle:2015:Shoal} showed that distribution
and replication of data mitigates congestion on interconnects and balances
memory controller load, by extending \gm~\cite{Hong:2012:GDE:2150976.2151013},
a high-level domain-specific language for graph analytics, to automatically
apply these techniques per region, using patterns extracted by the compiler.
This gave a two-fold speedup of already tuned parallel programs.

Large pages interact with the NUMA techniques described above, by changing the
granularity at which they can be applied to data structures that are
contiguous in virtual memory.  The granularity of NUMA distribution, for
example, is the page size. Hence, the smaller the page size the more slack the
run-time has to distribute data across NUMA nodes.  Bigger page sizes also
make memory allocation more restrictive: The starting address when allocating
memory must be a multiple of the page size.  Bigger page sizes can increase
fragmentation and increases the chance of conflicts in caches and TLB.

In \Sys, programs map their own memory, and all combinations of page sizes are
supported. Furthermore, no complex setup of page allocations and kernel
configurations are required.

Table~\ref{tab:pagerank_single-vs-multithreaded} shows the effect of the page
size on application performance using \shoal's
\gm \pagerank~\cite{Kaestle:2015:Shoal}.  NUMA effects are minimal on the 2-socket machine we are
using in other experiments, so for this experiment we use the machine in
Table~\ref{tab:sgs-r815-03} and note that AMD's SMT threads (CMT) are disabled
in our experiments.

\begin{table}[tb]
\center
\begin{tabular}{ll}
	%machine & \machinebig \\
	CPU                           & AMD Opteron 6378 \\
	micro architecture            & Piledriver \\
	\#nodes / \#sockets / \#cores & 8 / 4 / 32 @ 2.4 GHz \\
	L1 / L2 cache size            & \KB{16} / \MB{2} per core \\
	L3 cache size                 & \MB{12} per socket \\
	dTLB (\KB{4} pages)           & 64 entries, fully \\
	dTLB (\MB{2/4} pages)         & 64 entries, fully\\
	dTLB (\GB{1} pages)           & 64 entries, fully \\ % fully associative
	L2 TLB (\KB{4} pages)         & 1024 entries, 8 way \\
	L2 TLB (\MB{2/4} pages)       & 1024 entries, 8 way \\
	L2 TLB (\GB{1} pages)         & 1024 entries, 8 way  \\
	RAM                           & \GB{512} (\GB{64} per node) \\
\end{tabular}
\caption{Specification of machine used in \S{\ref{sec:eval:heteropages}}}
\label{tab:sgs-r815-03}
\end{table}

\begin{table}[t]
  \centering
  \begin{tabular}{l|rrr}
    page size & \multicolumn{3}{c}{array configuration} \\
     & T=1 & T=32 (dist) & T=32 (repl + dist) \\
    \hline
    \KB{4} & 597.91          & \textbf{51.32} &  34.43 \\
    \MB{2} & 414.80          &  58.09         & \textbf{28.87} \\
    \GB{1} & \textbf{395.64} & 265.94         & 128.77 \\
  \end{tabular}
  \caption{\pagerank runtime (seconds) depending on page size and
    \pagerank configuration (repl = replication,
    dist = distribution, T is the number of threads). Highlighted
    are best numbers for each configuration. Standard error is
    very small.}
  \label{tab:pagerank_single-vs-multithreaded}
\end{table}

We evaluate two configurations: First, single-threaded (T=1). In this case
replication does not make sense as all accesses are local, and distribution is
unnecessary as a single thread cannot saturate the memory controller ---
indeed, an increase in remote memory access would likely reduce performance.
In this case, an isolated application, bigger pages are always better.

Next, we run on all cores and explore the impact of replication and
distribution on the choice of page sizes.  \GB{1} pages clearly harm
performance as distribution is impossible or too coarse-grained. We only
break even if 90\% of the working set is replicated.  However, the last 10\%
still cannot be distributed efficiently, which leads to worse performance.

It is clear that the right page size is highly dynamic and depends on workload
and application characteristics. It is impractical to statically configure a
system with pools (as in Linux) optimally for all programs, as the
requirements are not known beforehand. Also, memory allocated to pools is not
available for allocations with different page sizes.  In contrast, \Sys's
simpler interface allows arbitrary use of page sizes and replication by the
application without requiring \emph{a priori} configuration of the OS.

\subsection{Page status bits}\label{sec:eval:boehm}

The potential of using the MMU to improve garbage collection is
known~\cite{Appel:1991:VMP:106972.106984}. Out of many possible applications,
we consider detecting page modifications; A feature used, for example, in the
Boehm garbage collector~\cite{Boehm:1991} to avoid stopping the world. Only
after tracing does the collector stop the world and perform a final trace that
need only consider marked objects in dirty pages. This way, newly reachable
objects are accounted for and not collected.

There are two ways to detect modified pages:  The first is to make the pages
read-only (e.g., via \pr{mprotect()} or transparently by the kernel using
soft-dirty PTEs~\cite{lx:soft_dirty}), and handle page faults in user-space
or kernel-space. The handler sets a virtual dirty bit, and unprotects the page
to allow the program to continue.  The second approach uses hardware dirty
bits, set when a page is updated. Some OSes (e.g., Linux) do not provide
access to these bits. This is not just an interface issue. The bits are
actively used by Linux to detect pages that need to be flushed to disk during
page reclamation. Other OSes such as Solaris expose these dirty bits in a
read-only manner via the \pr{/proc} file-system. In this case, applications
are required to perform a system call to read the bits, which, can lead to
worse performance than using \pr{mprotect()}~\cite{boehmgcdesc}.

In \Sys, physical memory and page tables are directly visible to applications.
Applications can map page tables read-only in their virtual address space.
Only clearing the dirty bits requires a system call.

Dune~\cite{Belay_dune} provides this functionality through nested paging
hardware, intended for virtualization, by running applications as a guest OS.
Dune applications have direct access to the virtualized (nested) page tables.
This approach avoids any system call overhead to reset the dirty bits, but
depends on virtualization hardware and can lead to a performance penalty due
to greater TLB usage~\cite{Basu:2013:EVM:2485922.2485943,Bhargava:2008}.

\begin{figure}[t]
    \begin{center}
        \includegraphics[page=1,width=7cm]{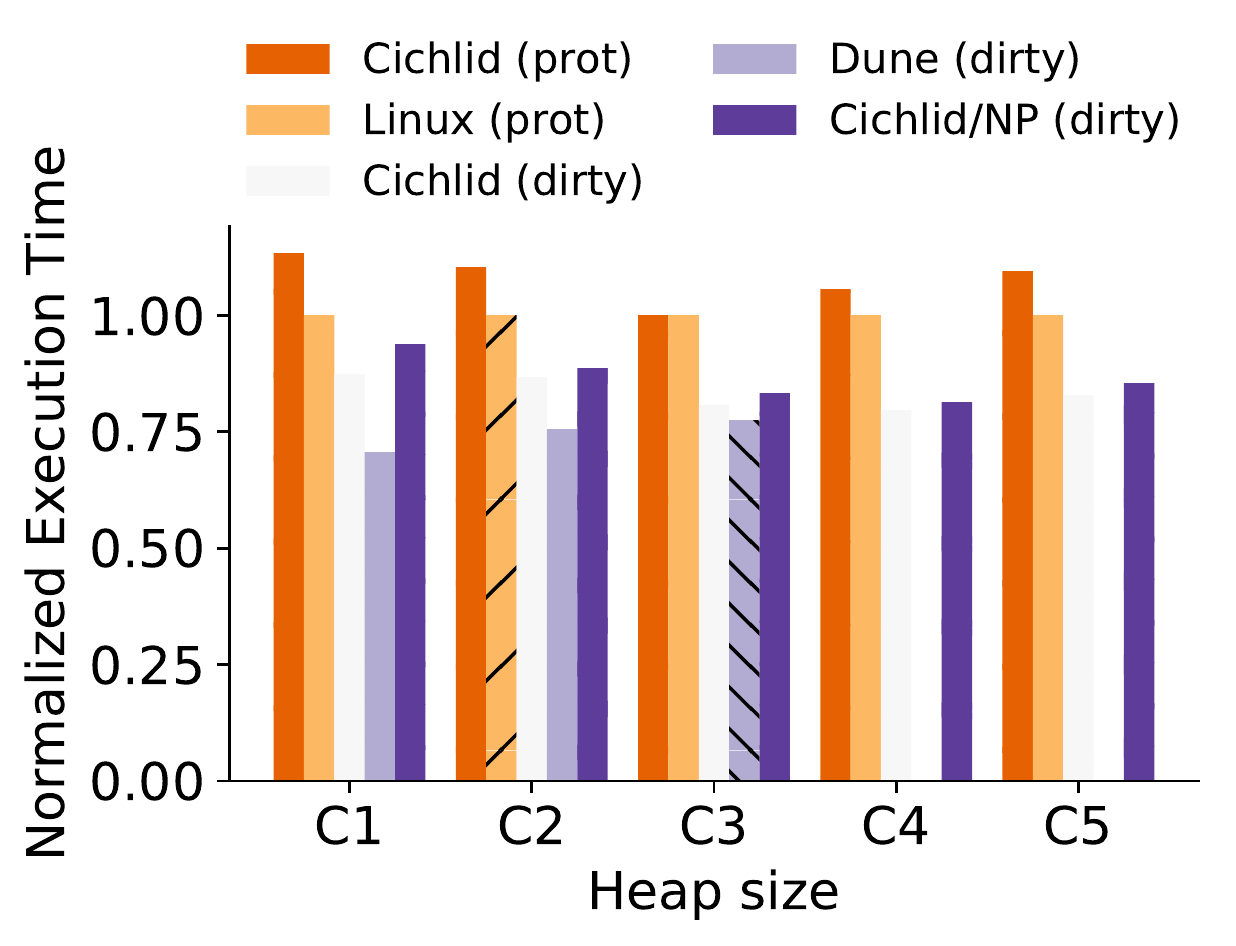}
    \end{center}
    \caption{GCBench on Linux, \Sys and Dune, normalized runtime to Linux.
        (Linux \linthree, \lindune)}
    \label{fig:gc-runtime}
\end{figure}

\begin{table}[t]
  \centering
  \begin{tabular}{lrrrrr}
    Config       & C1 & C2 & C3 & C4 & C5 \\
    \hline
    Runtime (s) \\
    \hline
    Linux (prot) & 2.1 & 9.6 & 42 & 191 & 848 \\
    \Sys (prot)  & 2.4 & 10.5& 43 & 203 & 928 \\
    \Sys (dirty) & 1.9 & 8.3 & 34 & 153 & 692 \\
    Dune (dirty) & 1.5 & 7.3 & 33 & -- & -- \\
    \Sys/NP (dirty) & 2.0 & 8.6 & 36 & 157 & 720 \\
    \hline
    Collections \\
    \hline
    Linux (prot) & 251 & 336 & 381 & 428 & 448 \\
    \Sys (prot)  & 245 & 335 & 393 & 432 & 442 \\
    \Sys (dirty) & 230 & 323 & 383 & 435 & 441 \\
    Dune (dirty) & 318 & 367 & 403 & -- & -- \\
    \Sys/NP (dirty) & 233 & 325 & 381 & 434 & 443 \\
    \hline
    Heap size (MB) \\
    \hline
    Linux (prot) & 139 & 411 & 1924 & 7972 & 24932 \\
    \Sys (prot)  & 132 & 453 & 1413 & 6789 & 26821 \\
    \Sys (dirty) & 100 & 453 & 1477 & 5669 & 28132 \\
    Dune (dirty) & 106 & 386 & 1579 & -- & -- \\
    \Sys/NP (dirty) & 100 & 453 & 1573 & 5541 & 28132 \\

  \end{tabular}
  \caption{GCBench reported total runtime, heap size and amount of collections.}
  \label{tab:GCBench Numbers}
  \label{tbl:gc}
\end{table}

We use the Boehm garbage collector~\cite{Boehm:1991} and the GCBench
microbenchmark~\cite{gcbench}. GCBench tests the garbage collector by
allocating and collecting binary trees of various sizes. We run this benchmark
with the three described memory systems, Linux, Dune and \Sys with five
different configurations C1 to C5, which progressively increase the size of the
allocated trees.

In Figure~\ref{fig:gc-runtime} we compare the runtime of each system. \Sys
implements all three mechanisms: protecting pages (\Sys (prot)), hardware
dirty bits (\Sys (dirty)) in user-space and hardware dirty bits in guest ring
0 (\Sys/NP (dirty)) (as does Dune). Our virtualization code is based on
Arrakis~\cite{peter_osdi_2014}. 

\Sys (prot) performs slightly worse than Linux (prot). This is
consistent with Figure~\ref{fig:results:memops-compare} where Linux
performs better than \Sys for protecting a single \KB{4} page.  We
achieve better performance (between 13\% (C2) and 19\% (C4)) than Linux
when we use hardware dirty bits, by avoiding traps when writing to
pages. We still incur some overhead as we have to make a system call to
reset the dirty bits on pages.  Dune outperforms \Sys (dirty) by up to
21\% (C1), as direct access to the guest page tables enables resetting
the dirty bits without having to make a system call.  However, \Sys
manages to close the gap as the working set becomes larger, in which
case Dune performance noticeably shows the overhead of nested paging.
Unfortunately, we were unable to get Dune working with larger heap sizes
on our hardware and thus have no numbers for Dune for configurations C4
and C5.

On Linux, using transparent huge pages did not have a significant impact
on performance and we report the Linux numbers with THP disabled.  In a
similar vein, we were unable to get Dune working with superpages, but we
believe that having superpages might improve Dune performance for larger
heap sizes (c.f.  \ref{sec:eval:gups}).

\Sys/NP (dirty) runs GCBench in guest ring 0 and reads and clears dirty
bits directly on the guest hardware page tables. The performance for
\Sys/NP is similar to \Sys (dirty) and slower than Dune.  However, this
can be attributed to the fact that \Sys/NP does not fully leverage the
advantage of having direct access to the guest hardware page tables and
still uses system calls to construct the address space.

Table~\ref{tbl:gc} shows the total runtime, number of collections the GC did
and the heap size used by the application. Ideally, the heap size should be
identical for all systems since it is always possible to trade memory for
better run time in a garbage collector. In practice this is very
difficult to enforce especially across entirely different operating
systems. For example \Sys uses less memory (28\%) for C4 compared to
Linux (prot) but more memory (12\%) for C5.

We conclude that with \Sys we can safely expose MMU information
to applications which in turn can benefit from it without relying on
virtualization hardware features.

%
% GUPS with nested paging (Dune)
% ----------------------------------------------------------------------------

\subsection{Nested paging overhead}\label{sec:eval:nestedgups}

To illustrate the potential downside of nested paging, we
revisit the HPC Challenge RandomAccess benchmark. Resolving a TLB miss with
nested paging requires a 2D page table walk and up to 24 memory
accesses~\cite{Ahn:2012:RHP:2337159.2337214} resulting in a much higher miss
penalty, and the overhead of nested paging may end up outweighing the benefits
of direct access to privileged hardware in guest ring zero.  GUPS represents a
worst-case scenario due to its lack of locality.

%\begin{figure}[t]
%	\begin{center}
%		\includegraphics[width=\columnwidth]{plots/gups-dune-absolute.pdf}
%	\end{center}
%	\caption{Comparison of the execution time of GUPS with and without
%    nested paging for varying working set sizes.}
%	\label{fig:results:gups-dune}
%\end{figure}

\begin{figure}[t]
	\begin{center}
		\includegraphics[width=\columnwidth]{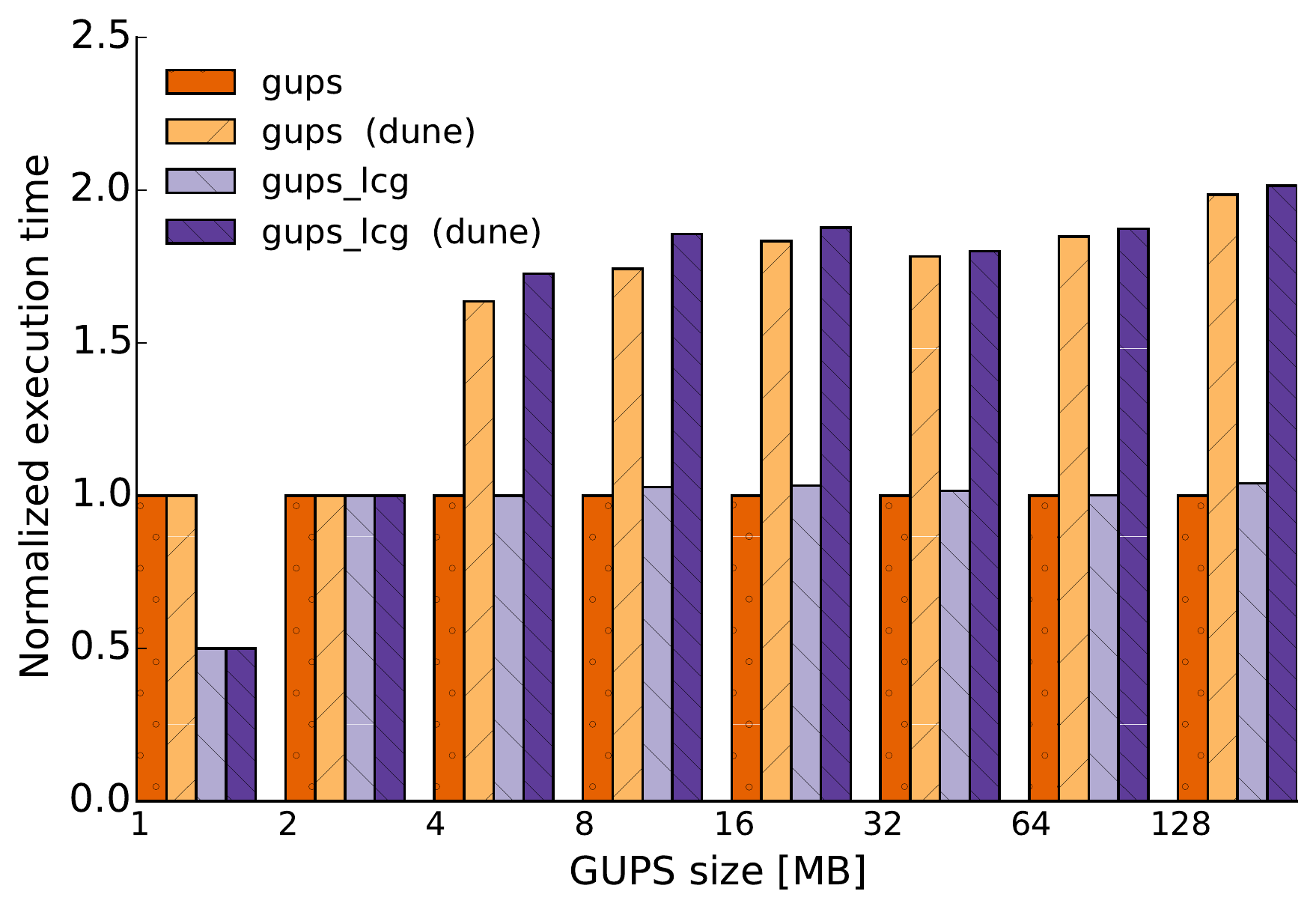}
	\end{center}
	\caption{Comparison of the execution time of RandomAccess with and without
    nested paging for varying working set sizes, normalized to GUPS on
    native Linux. (Linux \linthree, \lindune)}
	\label{fig:results:gups-dune-normalized}
\end{figure}

We conduct the same experiment as in section~\ref{sec:eval:gups} on
Dune~\cite{Belay_dune} with a working set size ranging from \MB{1} to
\MB{128}.  
Figure~\ref{fig:results:gups-dune-normalized} and Table~\ref{tbl:gupsvirtual}
show that for the smallest table sizes (\MB{1} and \MB{2}) the
performance of RandomAccess under Dune and Linux is comparable. Larger
working set sizes exceed the TLB coverage and hence more TLB misses occur.
This results in almost 2x higher runtime for RandomAccess in Dune than Linux.
As for all comparisons with Dune, we disable transparent huge pages
on Linux.

Running applications in guest ring zero as in Dune has pros and cons:
on one hand, the application gets access to privileged hardware
features, on the other hand, the performance may be degraded due to
larger TLB miss costs for working sets which cannot be covered by the
TLB.

\begin{table}[t]
  \centering
  \begin{tabular}{r|rr|rr}
         & Linux &  & Dune &   \\
    Size & GUPS & GUPS LCG & GUPS & GUPS LCG \\
    \hline
    1   & 2 & 1 & 2 & 1 \\
    2   & 3 & 3 & 3 & 3 \\
    4   & 11 & 11 & 18 & 19 \\
    8   & 35 & 36 & 61 & 65 \\
    16  & 90 & 93 & 165 & 169 \\
    32  & 236 & 240 & 421 & 425 \\
    64  & 594 & 595 & 1098 & 1113 \\
    128 & 1510 & 1571 & 2999 & 3043 \\
  \end{tabular}
  \caption{RandomAccess absolute execution times in milliseconds.
    (Linux \linthree, \lindune)}
  \label{tbl:gupsvirtual}
\end{table}

\subsection{Page coloring}

The core principle of paged virtual memory is that virtual pages are backed by
arbitrary physical pages. This can adversely affect application performance
due to unnecessary conflict misses in the CPU caches and an increase in
non-determinism~\cite{kim2011page}.  In addition, system wide page coloring
introduces constraints on memory management which may interfere with the
application's memory requirements~\cite{Zhang:2009:TPP}.

Implementing page placement policies is non-trivial: The complexity of
the FreeBSD kernel is increased
significantly~\cite{bsd_page_coloring}, Solaris allows applications to
chose from multiple algorithms~\cite{solaris_page_coloring}, and there
have been several failed attempts to implement page placement
algorithms in Linux. Other systems like COLORIS~\cite{Ye:2014:CDC}
replace Linux' page allocator entirely in order to support page
coloring.

% Application: Database Join. 
In contrast, \Sys allows an application to explicitly request physical memory of 
a certain color and map according to its needs. For instance, a streaming database join 
operator can restrict the large relation (which is streamed from disk) to a
small portion of the cache as most accesses would result in a cache miss
anyway and keep the smaller relation completely in cache. 

Table~\ref{tab:gups:color} shows the results of parallel execution of two
instances of the HPC Challenge suite RandomAccess benchmark on cores that
share the same last-level cache. In the first column we show the performance
of each instance running in isolation.  We see a significant drop in GUP/s for
the instance with the smaller working set when both instances run in parallel.
By applying cache partitioning we can keep the performance impact on the
smaller instance to a minimum while improving the performance of the
larger instance even compared to the case where the larger instance runs in
isolation.

The reason behind this unexpected performance improvement is that the working
set (the table) of the larger instance is restricted to a small fraction of
the cache which reduces conflict misses between the working set and other data
structures such as process state etc.

\begin{table}
    \begin{center}
    \begin{tabular}{l|c|cc|cc}
		Process & Isolation & 
            \multicolumn{2}{l|}{Parallel} & 
            \multicolumn{2}{l}{Parallel Colors} \\
        \hline
        16M Table & 0.0926 & 0.0834 & 90.0\% & 0.0921 & 99.5\% \\
        %          1.45s  & 1.61s & 1.46 & 
        64M Table & 0.0570 & 0.0561 & 98.4\% & 0.0631 & 110.7\% \\
        %          9.426s & 9.57s & 8.515
    \end{tabular}
    \caption{Parallel execution of GUPS on \Sys with and without cache 
    coloring. Values in GUP/s.}
    \label{tab:gups:color}
        \end{center}
\end{table}

\subsection{Discussion}\label{sec:eval:discussion}

With this evaluation, we have shown that the flexibility of \Sys's
memory system allows applications to optimize their physical resources
for a particular workload independent of a system-wide policy without
sacrificing performance.

\Sys's strength lies in its flexibility.  By stripping back the
policies baked into traditional VM systems over the years (many
motivated by RAM as a scarce resource) and exposing hardware resources
securely to programs, it performs as well as or better than Linux for
most benchmarks, while enabling performance optimizations not
previously possible in a clean manner.

%% file: evaluation-memops.tex
\subsection{Memory operation microbenchmarks}\label{sec:eval:micro}
\label{sec:eval:memops:compare}

\begin{figure*}[ht]
	\begin{center}
		\includegraphics[width=\textwidth]{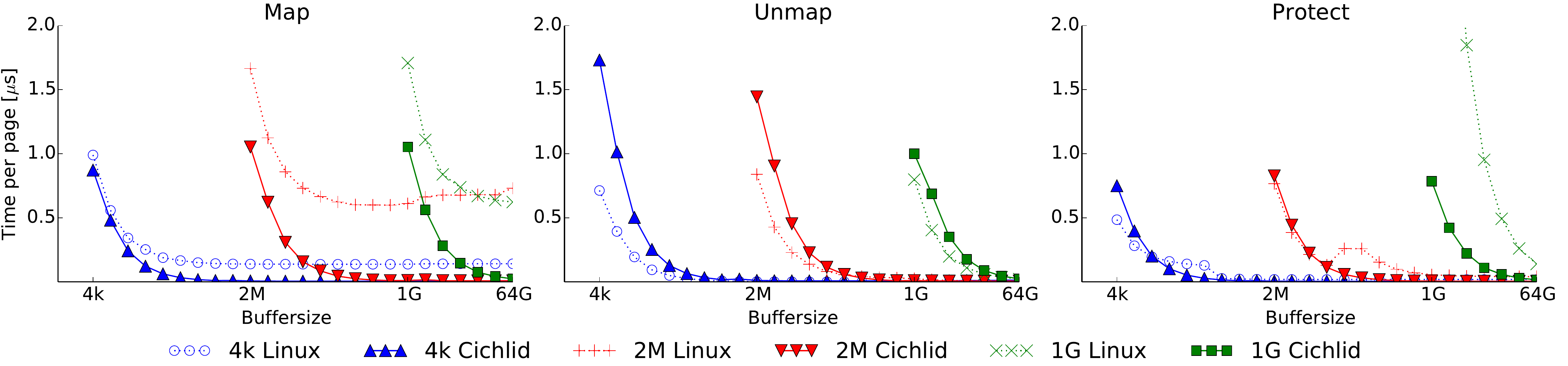}
	\end{center}
	\caption{Comparison of memory operations on \Sys and Linux using 
	\texttt{shmat}, \texttt{mprotect} and \texttt{shmdt}. (Linux \lintlb)}
	\label{fig:results:memops-compare}
\end{figure*}

%\paragraph{Experiment description}
We extend the Appel and Li benchmarks, to establish how the primitive
operations scale for large address spaces, using buffers up to \GB{64}. We
\emph{map}, \emph{protect} and \emph{unmap} the entire buffer, and time each
operation separately.  We compare \Sys to the best Linux method for each page
size established in \S~\ref{subsec:linux:discussion}. On \Sys we use the
high-level interfaces on a previously allocated frame, for similar semantics
to shared memory objects in Linux. The experiments were conducted on a
\babybel.  Figure~\ref{fig:results:memops-compare} shows execution time per
page.

%
% -----------------------------------------------------------------------------
%
%\subsubsection{Comparing Linux with \Sys}
\textbf{Map:} \Sys per-page performance is highly predictable,
regardless of page size.  Since all information needed is presented to
each a system call, the kernel does very little.  On Linux we use
\texttt{shm\_open} for 4k pages and \texttt{shmat} for others. Linux
needs to consult the shared segment descriptor and validate it.  This
results in a general performance improvement for \Sys over Linux up to 15x for
\KB{4} pages or 93x for large pages, once some upfront overhead is
amortized.

\textbf{Protect:} These are in line with the Appel and Li benchmarks: \Sys
outperforms Linux's \texttt{mprotect()} on an \texttt{mmap}'ed region in all
configurations except for small buffers of \KB{4} pages. For large buffers,
the differences between \Sys and Linux are up to 4x (\KB{4} pages) or 8x (huge
pages). 

\textbf{Unmap:} Doing an unmap in \Sys is expensive: the relevant page table
capability must be looked up to invoke it and the mapped \texttt{Frame}
capability needs to be marked unmapped.  Linux \texttt{shmdt}, however,
simply detaches the segment from the process but doesn't destroy it.  \Sys
could be modified to directly invoke the page table, and thereby match
the performance of Linux.

\Sys memory operations are competitive: capabilities and fast traps allows an
efficient virtual memory interface.  Even when multiple page table levels are
changed, \Sys usually outperforms Linux on most cases, despite requiring
several system calls.

%% file: related.tex
\section{Related work}\label{sec:related}

Prior to Barrelfish and seL4, the idea of moving memory management
into the application rather than a kernel or external paging server
had been around for some time. Engler et al. in
1995~\cite{Engler:1995:AAV:822074.822385} outlined 
much of the motivation for moving memory management into the
application rather than the kernel or external paging server, and
described AVM, an implementation for the
Exokernel~\cite{exokernel:sosp95} based on a software-loaded TLB,
presenting a small performance evaluation on microbenchmarks.  AVM
referred to physical memory explicitly by address, and ``secure
bindings'' conferred authorization to map it.  Since then,
software-loaded TLBs have fallen out of favor due to hardware
performance trends. \Sys targets modern hardware page tables, and uses
capabilities to both name and authorize physical memory access.

The V++ Cache Kernel~\cite{Cheriton:1994:CMO:1267638.1267652} implemented
user-level management of physical memory through page-frame caches
~\cite{Harty:1992:APM:143365.143511} allowing applications to monitor and 
control the physical frames they have, with a focus on better
page-replacement policies.  A virtual address space is a segment 
which is composed of regions from other segments called bound regions. 
A segment manager, associated with each segment, is responsible for keeping 
track of the segment to page mappings and hence handling page faults. Pages are 
migrated between segments to handle faults. Segment managers can run separately 
from the faulting application. It is critical to avoid double faults in the 
segment manager. Initialization is handled by the kernel which creates a 
well-known segment.	

Other systems have also reflected page faults to user space.
Microkernels like L4~\cite{Liedtke:1999:SUM:822076.822455},
Mach~\cite{Rashid:1988:MachVM},
Chorus~\cite{Abrossimov:1989:GVM:74850.74863}, and
Spring~\cite{Khalidi:1993:SpringVM} allow server processes to
implement custom page management policies.   
In contrast, the soft-realtime requirements of continuous media
motivated Nemesis~\cite{Hand:1999:SNO:296806.296812} redirecting
faults to the application itself, to ensures resource accountability.
As with AVM, the target hardware is a uniprocessor with a
software-loaded TLB.  A similar upcall mechanism for reflecting page
faults was used in K42~\cite{k42:eurosys06}.

In contrast, extensible kernels like
SPIN~\cite{Bershad:1995:ESP:224056.224077} and
VINO~\cite{Vino:1994:comp} allow downloading of safe policy extensions
into the kernel for performance.  For example, SPIN's kernel interface
to memory has some similarity with \Sys's user-space API:
\texttt{PhysAddr} allowed allocation, deallocation, and reclamation of
physical memory, \texttt{VirtAddr} managed a virtual address space,
and \texttt{Translation} allowed the installation of mappings between
the two, as well as event handlers to be installed for faults.  In
comparison, \Sys allows applications to define policies completely in
user-space, whereas SPIN has to rely on compiler support to make sure
the extensions are safe for use in kernel-space.

%% file: conclusion.tex
\section{Conclusion}\label{sec:conclusion}

\Sys inverts the classical VM model and securely exposes physical
memory and MMU hardware to applications without recourse
to virtualization hardware. It enables a variety of
optimizations based on the memory system which are either impossible
to express in Unix-like systems, or can only be cast as ``hints'' to a
fixed kernel policy.  Although MMU hardware has evolved to support
a Unix-oriented view of virtual memory, \Sys outperforms the Linux VM
in many cases, and equals it in others. 

\Sys explores a very different style of OS service provision.  Demand
paging often badly impacts modern applications that
rely on fast memory; the virtual address space can be an abstraction
barrier which degrades performance.  In \Sys, in contrast, an
application \emph{knows} when it has insufficient physical memory and
must explicitly deal with it.  Given current trends in both
applications and hardware, we feel this ``road less travelled'' in OS
design is worthy of further attention. 
Exposing hardware securely to applications, libraries, and language
runtimes may be the only practical way to avoid the increasing
complexity of memory interfaces based purely on virtual addressing.

%% file: main.bbl
\providecommand{\noopsort}[1]{} \providecommand{\url}{\error{The bib files now
  require `url' package!}}
\begin{thebibliography}{10}

\bibitem{Abrossimov:1989:GVM:74850.74863}
{\sc Abrossimov, E., Rozier, M., and Shapiro, M.}
\newblock Generic {V}irtual {M}emory {M}anagement for {O}perating {S}ystem
  {K}ernels.
\newblock In {\em Proceedings of the Twelfth ACM Symposium on Operating Systems
  Principles\/} (1989), SOSP '89, ACM, pp.~123--136.

\bibitem{Ahn:2012:RHP:2337159.2337214}
{\sc Ahn, J., Jin, S., and Huh, J.}
\newblock {R}evisiting {H}ardware-assisted {P}age {W}alks for {V}irtualized
  {S}ystems.
\newblock In {\em Proceedings of the 39th Annual International Symposium on
  Computer Architecture\/} (Washington, DC, USA, 2012), ISCA '12, IEEE Computer
  Society, pp.~476--487.

\bibitem{Appel:1991:VMP:106972.106984}
{\sc Appel, A.~W., and Li, K.}
\newblock Virtual {M}emory {P}rimitives for {U}ser {P}rograms.
\newblock In {\em Proceedings of the Fourth International Conference on
  Architectural Support for Programming Languages and Operating Systems\/} (New
  York, NY, USA, 1991), ASPLOS IV, ACM, pp.~96--107.

\bibitem{armv7}
{\sc {ARM Ltd.}}
\newblock {\em Cortex-{A}9 {T}echnical {R}eference {M}anual}.
\newblock Revision r4p1.

\bibitem{oracle:thp2}
{\sc Aziz, K.}
\newblock {I}mproving the {P}erformance of {T}ransparent {H}uge {P}ages in
  {L}inux.
\newblock
  \url{https://blogs.oracle.com/linuxkernel/entry/performance_impact_of_transparent_huge},
  Aug 2014.

\bibitem{Barr:2010:TCS:1815961.1815970}
{\sc Barr, T.~W., Cox, A.~L., and Rixner, S.}
\newblock Translation {C}aching: {S}kip, {D}on't {W}alk (the {P}age {T}able).
\newblock In {\em Proceedings of the 37th Annual International Symposium on
  Computer Architecture\/} (New York, NY, USA, 2010), ISCA '10, ACM,
  pp.~48--59.

\bibitem{Basu:2013:EVM:2485922.2485943}
{\sc Basu, A., Gandhi, J., Chang, J., Hill, M.~D., and Swift, M.~M.}
\newblock {E}fficient {V}irtual {M}emory for {B}ig {M}emory {S}ervers.
\newblock In {\em Proceedings of the 40th Annual International Symposium on
  Computer Architecture\/} (New York, NY, USA, 2013), ISCA '13, ACM,
  pp.~237--248.

\bibitem{barrelfish}
{\sc Baumann, A., Barham, P., Dagand, P.-E., Harris, T., Isaacs, R., Peter, S.,
  Roscoe, T., Sch\"{u}pbach, A., and Singhania, A.}
\newblock The {M}ultikernel: a new {OS} architecture for scalable multicore
  systems.
\newblock In {\em Proceedings of the 22nd ACM Symposium on Operating Systems
  Principles\/} (2009), pp.~29--44.

\bibitem{Belay_dune}
{\sc Belay, A., Bittau, A., Mashtizadeh, A., Terei, D., Mazi\`{e}res, D., and
  Kozyrakis, C.}
\newblock Dune: safe user-level access to privileged {CPU} features.
\newblock In {\em Proceedings of the 10th USENIX conference on Operating
  Systems Design and Implementation (OSDI)\/} (Hollywood, CA, USA, 2012).

\bibitem{Bershad:1995:ESP:224056.224077}
{\sc Bershad, B.~N., Savage, S., Pardyak, P., Sirer, E.~G., Fiuczynski, M.~E.,
  Becker, D., Chambers, C., and Eggers, S.}
\newblock Extensibility {S}afety and {P}erformance in the {SPIN} {O}perating
  {S}ystem.
\newblock In {\em Proceedings of the Fifteenth ACM Symposium on Operating
  Systems Principles\/} (New York, NY, USA, 1995), SOSP '95, ACM, pp.~267--283.

\bibitem{Bhargava:2008}
{\sc Bhargava, R., Serebrin, B., Spadini, F., and Manne, S.}
\newblock Accelerating {T}wo-dimensional {P}age {W}alks for {V}irtualized
  {S}ystems.
\newblock In {\em Proceedings of the 13th International Conference on
  Architectural Support for Programming Languages and Operating Systems\/}
  (2008), ASPLOS XIII, pp.~26--35.

\bibitem{Bhattacharjee:2013:LMM:2540708.2540741}
{\sc Bhattacharjee, A.}
\newblock Large-reach {M}emory {M}anagement {U}nit {C}aches.
\newblock In {\em Proceedings of the 46th Annual IEEE/ACM International
  Symposium on Microarchitecture\/} (New York, NY, USA, 2013), MICRO-46, ACM,
  pp.~383--394.

\bibitem{boehmgcdesc}
{\sc Boehm, H.-J.}
\newblock Conservative {GC} algorithmic overview.
\newblock \url{http://www.hboehm.info/gc/gcdescr.html}.

\bibitem{gcbench}
{\sc Boehm, H.-J.}
\newblock Gcbench.
\newblock \url{http://hboehm.info/gc/gc_bench/}.

\bibitem{Boehm:1991}
{\sc Boehm, H.-J., Demers, A.~J., and Shenker, S.}
\newblock Mostly {P}arallel {G}arbage {C}ollection.
\newblock In {\em Proceedings of the ACM SIGPLAN 1991 Conference on Programming
  Language Design and Implementation\/} (1991), PLDI '91, pp.~157--164.

\bibitem{Boyd-Wickizer:2008:COS:1855741.1855745}
{\sc Boyd-Wickizer, S., Chen, H., Chen, R., Mao, Y., Kaashoek, F., Morris, R.,
  Pesterev, A., Stein, L., Wu, M., Dai, Y., Zhang, Y., and Zhang, Z.}
\newblock {C}orey: {A}n {O}perating {S}ystem for {M}any {C}ores.
\newblock In {\em Proceedings of the 8th USENIX Conference on Operating Systems
  Design and Implementation\/} (Berkeley, CA, USA, 2008), OSDI'08, USENIX
  Association, pp.~43--57.

\bibitem{oracle:thp}
{\sc Casey, M.}
\newblock Performance {I}ssues with {T}ransparent {H}uge {P}ages ({THP}).
\newblock
  \url{https://blogs.oracle.com/linux/entry/performance_issues_with_transparent_huge},
  Sep 2013.

\bibitem{Cheriton:1994:CMO:1267638.1267652}
{\sc Cheriton, D.~R., and Duda, K.~J.}
\newblock A {C}aching {M}odel of {O}perating {S}ystem {K}ernel {F}unctionality.
\newblock In {\em Proceedings of the 1st USENIX Conference on Operating Systems
  Design and Implementation\/} (Monterey, California, 1994), OSDI '94, USENIX
  Association.

\bibitem{lwn:autonuma12}
{\sc Corbet, J.}
\newblock {AutoNUMA}: the other approach to {NUMA} scheduling.
\newblock \url{http://lwn.net/Articles/488709/}, Mar 2012.

\bibitem{lwn:numahurry12}
{\sc Corbet, J.}
\newblock {NUMA} in a hurry.
\newblock \url{http://lwn.net/Articles/524977/}, Nov 2012.

\bibitem{lwn:numasched12}
{\sc Corbet, J.}
\newblock Toward better {NUMA} scheduling.
\newblock \url{http://lwn.net/Articles/486858/}, Mar 2012.

\bibitem{lwn:numasched13}
{\sc Corbet, J.}
\newblock {NUMA} scheduling progress.
\newblock \url{http://lwn.net/Articles/568870/}, Oct 2013.

\bibitem{lwn:userfault13}
{\sc Corbet, J.}
\newblock User-space page fault handling.
\newblock \url{http://lwn.net/Articles/550555/}, May 2013.

\bibitem{lwn:thp_issues}
{\sc Corbet, J.}
\newblock 2014 {LSFMM} summit: Huge page issues.
\newblock \url{http://lwn.net/Articles/592011/}, Mar 2014.

\bibitem{lwn:numaschedprobs14}
{\sc Corbet, J.}
\newblock {NUMA} placement problems.
\newblock \url{http://lwn.net/Articles/591995/}, Mar 2014.

\bibitem{lwn:userfault14}
{\sc Corbet, J.}
\newblock Page faults in user space: {MADV\_USERFAULT}, remap\_anon\_range(),
  and userfaultfd().
\newblock \url{http://lwn.net/Articles/615086/}, Oct 2014.

\bibitem{lwn:transhuge}
{\sc Corbet, J.}
\newblock Transparent huge pages in 2.6.38.
\newblock \url{http://lwn.net/Articles/423584/}, Jan 2014.

\bibitem{dagand:fof:plos09}
{\sc Dagand, P.-E., Baumann, A., and Roscoe, T.}
\newblock {Filet-o-Fish}: practical and dependable domain-specific languages
  for {OS} development.
\newblock In {\em 5th Workshop on Programming Languages and Operating Systems
  (PLOS)\/} (Oct 2009).

\bibitem{Dashti:2013:TMH:2451116.2451157}
{\sc Dashti, M., Fedorova, A., Funston, J., Gaud, F., Lachaize, R., Lepers, B.,
  Quema, V., and Roth, M.}
\newblock {Traffic Management: A Holistic Approach to Memory Placement on NUMA
  Systems}.
\newblock In {\em Proceedings of the Eighteenth International Conference on
  Architectural Support for Programming Languages and Operating Systems\/}
  (Houston, Texas, USA, 2013), ASPLOS '13, ACM, pp.~381--394.

\bibitem{sel4:refman}
{\sc Derrin, P., Elkaduwe, D., and Elphinstone, K.}
\newblock {\em {seL4} {R}eference {M}anual}.
\newblock NICTA, 2006.
\newblock \url{http://www.ertos.nicta.com.au/research/sel4/sel4-refman.pdf}.

\bibitem{bsd_page_coloring}
{\sc Dillon, M.}
\newblock {Design elements of the FreeBSD VM system - Page Coloring}.
\newblock Online,
  \url{https://www.freebsd.org/doc/en/articles/vm-design/page-coloring-optimizations.html},
  Nov 2013.
\newblock Accessed 2015-08-26.

\bibitem{sel4-mm:mikes07}
{\sc Elkaduwe, D., Derrin, P., and Elphinstone, K.}
\newblock A memory allocation model for an embedded microkernel.
\newblock In {\em Proceedings of the 1st International Workshop on Microkernels
  for Embedded Systems (MIKES)\/} (2007), pp.~28--34.

\bibitem{sel4:iies08}
{\sc Elkaduwe, D., Derrin, P., and Elphinstone, K.}
\newblock Kernel {D}esign for {I}solation and {A}ssurance of {P}hysical
  {M}emory.
\newblock In {\em Proceedings of the 1st Workshop on Isolation and Integration
  in Embedded Systems\/} (New York, NY, USA, 2008), IIES '08, ACM, pp.~35--40.

\bibitem{Vino:1994:comp}
{\sc Endo, Y., Seltzer, M., Gwertzman, J., Small, C., Smith, K.~A., and Tang,
  D.}
\newblock {VINO}: {T}he 1994 {F}all {H}arvest.
\newblock Technical Report TR-34-94, Center for Research in Computing
  Technology, Harvard University, December 1994.

\bibitem{Engler:1995:AAV:822074.822385}
{\sc Engler, D.~R., Gupta, S.~K., and Kaashoek, M.~F.}
\newblock {AVM}: {A}pplication-level {V}irtual {M}emory.
\newblock In {\em Proceedings of the Fifth Workshop on Hot Topics in Operating
  Systems (HotOS-V)\/} (1995), HOTOS '95, IEEE Computer Society, pp.~72--.

\bibitem{exokernel:sosp95}
{\sc Engler, D.~R., Kaashoek, M.~F., and O'Toole, Jr., J.}
\newblock {Exokernel: An Operating System Architecture for Application-level
  Resource Management}.
\newblock In {\em Proceedings of the 15th ACM Symposium on Operating Systems
  Principles\/} (1995), pp.~251--266.

\bibitem{jemalloc:thp}
{\sc Evans, J.}
\newblock Issue \#243: Improve interaction with transparent huge pages.
\newblock \url{https://github.com/jemalloc/jemalloc/issues/243}, Jul 2015.

\bibitem{Gaud:2014:LPM:2643634.2643659}
{\sc Gaud, F., Lepers, B., Decouchant, J., Funston, J., Fedorova, A., and
  Qu{\'e}ma, V.}
\newblock {Large Pages May Be Harmful on NUMA Systems}.
\newblock In {\em Proceedings of the 2014 USENIX Conference on USENIX Annual
  Technical Conference\/} (Philadelphia, PA, 2014), USENIX ATC'14, USENIX
  Association, pp.~231--242.

\bibitem{Giceva:2014:DQP:2735508.2735513}
{\sc Giceva, J., Alonso, G., Roscoe, T., and Harris, T.}
\newblock Deployment of query plans on multicores.
\newblock {\em Proc. VLDB Endow. 8}, 3 (Nov 2014), 233--244.

\bibitem{lwn:hugepages}
{\sc Gorman, M.}
\newblock Huge pages.
\newblock \url{http://lwn.net/Articles/374424/}, Feb 2010.

\bibitem{lwn:libhugetlbfs}
{\sc Gorman, M.}
\newblock Huge pages part 2: Interfaces.
\newblock \url{https://lwn.net/Articles/375096/}, Feb 2010.

\bibitem{Gorman:2010:PCE:2185870.2185899}
{\sc Gorman, M., and Healy, P.}
\newblock Performance {C}haracteristics of {E}xplicit {S}uperpage {S}upport.
\newblock In {\em Proceedings of the 2010 International Conference on Computer
  Architecture\/} (Berlin, Heidelberg, 2012), ISCA'10, Springer-Verlag,
  pp.~293--310.

\bibitem{web:libsigsegv}
{\sc Haible, B., and Bonzini, P.}
\newblock {GNU} libsigsegv - {H}andling page faults in user mode.
\newblock \url{http://libsigsegv.sourceforge.net/}.

\bibitem{Hand:1999:SNO:296806.296812}
{\sc Hand, S.~M.}
\newblock Self-paging in the {N}emesis {O}perating {S}ystem.
\newblock In {\em Proceedings of the Third Symposium on Operating Systems
  Design and Implementation\/} (New Orleans, Louisiana, USA, 1999), OSDI '99,
  USENIX Association, pp.~73--86.

\bibitem{lx:x86tlbflush}
{\sc Hansen, D.}
\newblock {TLB} flushing on x86.
\newblock \url{https://www.kernel.org/doc/Documentation/x86/tlb.txt}.

\bibitem{Harty:1992:APM:143365.143511}
{\sc Harty, K., and Cheriton, D.~R.}
\newblock Application-controlled {P}hysical {M}emory {U}sing {E}xternal
  {P}age-cache {M}anagement.
\newblock In {\em Proceedings of the Fifth International Conference on
  Architectural Support for Programming Languages and Operating Systems\/} (New
  York, NY, USA, 1992), ASPLOS V, ACM, pp.~187--197.

\bibitem{Hong:2012:GDE:2150976.2151013}
{\sc Hong, S., Chafi, H., Sedlar, E., and Olukotun, K.}
\newblock {G}reen-{M}arl: {A} {DSL} for {E}asy and {E}fficient {G}raph
  {A}nalysis.
\newblock In {\em Proceedings of the Seventeenth International Conference on
  Architectural Support for Programming Languages and Operating Systems\/} (New
  York, NY, USA, 2012), ASPLOS XVII, ACM, pp.~349--362.

\bibitem{themachine_website}
{\sc {HP Labs}}.
\newblock The {M}achine.
\newblock \url{http://www.hpl.hp.com/research/systems-research/themachine/},
  January 2015.

\bibitem{intel:optref}
{\sc {Intel Corporation}}.
\newblock {\em Intel 64 and {IA-32} {A}rchitectures {O}ptimization {R}eference
  {M}anual}, September 2014.
\newblock Online. Accessed 2015-03-12.
  \url{http://www.intel.com/content/www/us/en/architecture-and-technology/64-ia-32-architectures-optimization-manual.html?wapkw=order+number+248966-025}.

\bibitem{Kaestle:2015:Shoal}
{\sc Kaestle, S., Achermann, R., Roscoe, T., and Harris, T.}
\newblock {Shoal: Smart Allocation and Replication of Memory for Parallel
  Programs}.
\newblock In {\em Proceedings of the 2015 USENIX Annual Technical Conference\/}
  (Santa Clara, CA, 2015), USENIX ATC '15, pp.~263--276.

\bibitem{Khalidi:1993:SpringVM}
{\sc Khalidi, Y.~A., and Nelson, M.~N.}
\newblock The {S}pring {V}irtual {M}emory {S}ystem.
\newblock Technical Report SMLI TR-93-9, Sun Microsystems Laboratories Inc.,
  February 1993.

\bibitem{kim2011page}
{\sc Kim, J., Kim, J., Ahn, D., and Eom, Y.~I.}
\newblock Page coloring synchronization for improving cache performance in
  virtualization environment.
\newblock In {\em Computational Science and Its Applications-ICCSA 2011}.
  Springer, 2011, pp.~495--505.

\bibitem{sel4:sosp09}
{\sc Klein, G., Elphinstone, K., Heiser, G., Andronick, J., Cock, D., Derrin,
  P., Elkaduwe, D., Engelhardt, K., Kolanski, R., Norrish, M., Sewell, T.,
  Tuch, H., and Winwood, S.}
\newblock {seL4}: Formal verification of an {OS} kernel.
\newblock In {\em Proceedings of the 22nd ACM Symposium on Operating Systems
  Principles\/} (2009).

\bibitem{gups_randomaccess}
{\sc Koester, D., and Lucas, B.}
\newblock {HPC} {C}hallenge - {R}andom {A}ccess.
\newblock Online.
\newblock \url{http://icl.cs.utk.edu/projectsfiles/hpcc/RandomAccess/}.
  Accessed 2015-03-09.

\bibitem{k42:eurosys06}
{\sc Krieger, O., Auslander, M., Rosenburg, B., Wisniewski, R.~W., Xenidis, J.,
  Da~Silva, D., Ostrowski, M., Appavoo, J., Butrico, M., Mergen, M., Waterland,
  A., and Uhlig, V.}
\newblock K42: {B}uilding a {C}omplete {O}perating {S}ystem.
\newblock In {\em Proceedings of the 1st EuroSys Conference\/} (2006),
  pp.~133--145.

\bibitem{Leis:2014:MPN:2588555.2610507}
{\sc Leis, V., Boncz, P., Kemper, A., and Neumann, T.}
\newblock Morsel-driven parallelism: A numa-aware query evaluation framework
  for the many-core age.
\newblock In {\em Proceedings of the 2014 ACM SIGMOD International Conference
  on Management of Data\/} (New York, NY, USA, 2014), SIGMOD '14, ACM,
  pp.~743--754.

\bibitem{Liedtke:1999:SUM:822076.822455}
{\sc Liedtke, J., Uhlig, V., Elphinstone, K., Jaeger, T., and Park, Y.}
\newblock How to {S}chedule {U}nlimited {M}emory {P}inning of {U}ntrusted
  {P}rocesses or {P}rovisional {I}deas {A}bout {S}ervice-{N}eutrality.
\newblock In {\em Proceedings of the The Seventh Workshop on Hot Topics in
  Operating Systems\/} (Washington, DC, USA, 1999), HOTOS '99, IEEE Computer
  Society, pp.~153--.

\bibitem{lx:hugetlbpage}
{\sc {Linux Kernel Project}}.
\newblock Hugetlbpage support in the {Linux} kernel.
\newblock \url{https://www.kernel.org/doc/Documentation/vm/hugetlbpage.txt}.

\bibitem{lx:transhuge}
{\sc {Linux Kernel Project}}.
\newblock Transparent {H}ugepage {S}upport.
\newblock \url{https://www.kernel.org/doc/Documentation/vm/transhuge.txt}.

\bibitem{Navarro:2002:PTO:1060289.1060299}
{\sc Navarro, J., Iyer, S., Druschel, P., and Cox, A.}
\newblock Practical, transparent operating system support for superpages.
\newblock {\em SIGOPS Oper. Syst. Rev. 36}, SI (Dec 2002), 89--104.

\bibitem{solaris_page_coloring}
{\sc {Oracle Corporation}}.
\newblock Online.
  \url{http://docs.oracle.com/cd/E19683-01/806-7009/chapter2-95/index.html},
  2010.
\newblock Accessed 2015-08-15.

\bibitem{hotos:themachine}
{\sc Paolo~Faraboschi, Kimberly~Keeton, T.~M., and Milojicic, D.}
\newblock Beyond processor-centric operating systems.
\newblock In {\em Proceedings of the 2015 International Workshop on Hot Topics
  in Operating Systems (HotOS XV)\/} (Karthause Ittingen, Warth-Weiningen,
  Switzerland, May 2015).

\bibitem{peter_osdi_2014}
{\sc Peter, S., Li, J., Zhang, I., Ports, D. R.~K., Woos, D., Krishnamurthy,
  A., Anderson, T., and Roscoe, T.}
\newblock {Arrakis: The Operating System is the Control Plane}.
\newblock In {\em 11th Symposium on Operating Systems Design and Implementation
  (OSDI'14)\/} (Broomfield, Colorado, USA, October 2014).

\bibitem{Rashid:1988:MachVM}
{\sc Rashid, R., Tevanian, A., J., Young, M., Golub, D., Baron, R., Black, D.,
  Bolosky, W., and Chew, J.}
\newblock Machine-{I}independent {V}irtual {M}emory {M}anagement for {P}aged
  {U}niprocessor and {M}ultiprocessor {A}rchitectures.
\newblock {\em Computers, IEEE Transactions on 37}, 8 (Aug 1988), 896--908.

\bibitem{redis:thp}
{\sc Sanfilippo, S.}
\newblock Redis latency problems troubleshooting.
\newblock \url{http://redis.io/topics/latency}.

\bibitem{lx:soft_dirty}
{S}oft-{D}irty {PTE}s.
\newblock \url{https://www.kernel.org/doc/Documentation/vm/soft-dirty.txt}.

\bibitem{Soma:2014:RVM:2612262.2612264}
{\sc Soma, Y., Gerofi, B., and Ishikawa, Y.}
\newblock Revisiting {V}irtual {M}emory for {H}igh {P}erformance {C}omputing on
  {M}anycore {A}rchitectures: {A} {H}ybrid {S}egmentation {K}ernel {A}pproach.
\newblock In {\em Proceedings of the 4th International Workshop on Runtime and
  Operating Systems for Supercomputers\/} (New York, NY, USA, 2014), ROSS '14,
  ACM, pp.~3:1--3:8.

\bibitem{gups}
{\sc {The University of Tennessee}}.
\newblock {HPC} {C}hallenge {B}enchmark.
\newblock Online.
\newblock \url{http://icl.cs.utk.edu/hpcc/software/view.html?id=178}. Accessed
  2015-03-09.

\bibitem{Ye:2014:CDC}
{\sc Ye, Y., West, R., Cheng, Z., and Li, Y.}
\newblock {COLORIS: A Dynamic Cache Partitioning System Using Page Coloring}.
\newblock In {\em Proceedings of the 23rd International Conference on Parallel
  Architectures and Compilation\/} (Edmonton, AB, Canada, 2014), PACT '14,
  pp.~381--392.

\bibitem{Zhang:2009:TPP}
{\sc Zhang, X., Dwarkadas, S., and Shen, K.}
\newblock {Towards Practical Page Coloring-based Multicore Cache Management}.
\newblock In {\em Proceedings of the 4th ACM European Conference on Computer
  Systems\/} (Nuremberg, Germany, 2009), EuroSys '09, pp.~89--102.

\end{thebibliography}
